\documentclass[12pt]{article}
\usepackage[utf8]{inputenc}
\usepackage[a4paper, top=0.45in, bottom=0.75in, left=0.75in, right=0.75in]{geometry}

\usepackage{authblk}
\usepackage{authblk}
\usepackage{soul} 
\usepackage{graphicx}
\usepackage{epstopdf, epsfig}    
\usepackage{amssymb}
\usepackage{float}
\usepackage{xcolor}
\usepackage{hyperref}
\usepackage{lscape}
\usepackage{newtxtext}
\usepackage{newtxmath}
\usepackage{lineno}
\usepackage{setspace}
\usepackage[normalem]{ulem}
\usepackage[authoryear, round]{natbib}
\setcitestyle{comma, round}
\usepackage{amsmath}
\numberwithin{equation}{section}
\usepackage{upgreek}
\usepackage{caption}
\usepackage{natbib}
\usepackage{doi}
\usepackage{titlesec}
\titlelabel{\thetitle. } 

\usepackage{booktabs}
\usepackage{pifont}

\usepackage{xcolor}
\providecommand{\keywords}[1]{\textbf{\textit{Keywords:}} #1}
\titleformat{\paragraph}[block]{\fseries}{\theparagraph}{1em}{}

\hypersetup{
    colorlinks = true,
    urlcolor   = blue,
    citecolor  = blue,
    linkcolor  = blue
}

\title{High-energy droplet collisions in multi-interacting hollow cone sprays}

\author[1]{Narendra Dev\thanks{Current affiliation: Univ Claude Bernard Lyon 1, LMFA, UMR5509, 69622, Villeurbanne, France.}\thanks{Corresponding author: \href{mailto:narendradev@alum.iisc.ac.in}{narendradev@alum.iisc.ac.in}}}
\author[2]{Varun Kulkarni}
\author[1]{Sivakumar Deivandren}

\affil[1]{Department of Aerospace Engineering, Indian Institute of Science, Bangalore 560012, India}
\affil[2]{School of Engineering and Applied Sciences, Harvard University, Cambridge, MA 02138, USA}
\date{}
\begin{document}

\maketitle

\begin{abstract}
Droplets collide in several complex spray environments ranging from sea sprays to combustion chambers, altering their size and velocity characteristics. The present work offers a systematic investigation of such collisions within the interacting region formed by three hollow-cone sprays, termed the combined spray, at two elevated liquid sheet Weber numbers ($W\!e_l$). The integrated analysis employs Phase Doppler Interferometry (PDI) and microscopic high-speed backlight imaging to characterize the collision dynamics. PDI indicates a notable reduction (11–15\%) in Sauter Mean Diameter (SMD) at the onset of the interaction region. Images reveal frequent and high-energy droplet collisions, capturing structures associated with binary collision outcomes, namely \textit{reflexive and stretching separations}, \textit{splashing}, \textit{fingering}, and \textit{stretching with digitations}, along with complex \textit{multi-droplet collisions}. These collisions produce numerous smaller satellite droplets at the expense of larger parent droplets, leading to a decrease in local SMD. Increasing $W\!e_l$ elevates the frequency of these outcomes, particularly highlighting \textit{stretching separation} as the dominant mechanism. Furthermore, joint probability density functions from PDI and image-based analysis confirm that most satellite droplets predominantly exhibit axial motion, in contrast to the initial trajectories of parent droplets. The satellite droplets continue to move downstream, colliding with others, resulting in a cascade effect, producing finer droplets. Rescaled droplet size distributions, normalised by mean droplet diameter, are broader in the combined spray due to enhanced size reduction from collisions. These distributions are well captured by the compound gamma distribution, reflecting ligament-mediated breakup dynamics.
\end{abstract}

\keywords{Atomization, Droplet collision, Droplet breakup, Phase Doppler Interferometry, High-speed microscopy}
\maketitle

\section*{Nomenclature}
\begingroup
\renewcommand{\thelinenumber}{}%
\renewcommand{\makeLineNumber}{}%
\begin{center}
\fcolorbox{black!70}{white}{%
\begin{minipage}{0.98\linewidth}
\renewcommand{\arraystretch}{0.7}  
\setlength{\tabcolsep}{5pt}       
\begin{tabular}{p{2.8cm}p{12cm}}

\multicolumn{2}{l}{\textbf{Variables}}\\
$d$ & Instantaneous droplet diameter ($\mu$m) \\
$d_s$ & Satellite droplet diameter measured using images ($\mu$m) \\
$G$ & Spacing between the nozzles \\
$U$, $V$, $W$ & Velocity components of droplets measured using PDI (m/s) \\
$v_r$ & Relative velocity between colliding droplets (m/s) \\

$U_s$, $V_s$ & Velocity components of satellite droplets measured using images (m/s) \\
$V_{res}$ & Resultant velocity of satellite measured using images \\
$W\!e$ & Collision Weber number \\

$W\!e_s$ & Symmetric collision Weber number \\
$W\!e_l$ & Liquid-sheet Weber number \\
$z$ & Axial location from nozzle orifice (mm) \\
$z/G$ & Normalized axial location \\[6pt]

\multicolumn{2}{l}{\textbf{Abbreviations}}\\
AMD & Arithmetic mean diameter calculated using PDI data ($\mu$m) \\
CS & Combined spray \\
JPDF & Joint probability density function \\
LDM & Long distance microscope\\
P1, P2 & Colliding parent droplets in images \\
PDI & Phase Doppler Interferometry \\
SMD & Sauter mean diameter calculated using PDI data ($\mu$m) \\
SS & Single spray \\
TC & Taylor–Culick rim \\

\end{tabular}
\end{minipage}}
\end{center}
\endgroup

\section{Introduction}
\label{sec:intro}

\hspace{0.65cm}Atomization is a ubiquitous process observed in natural phenomena such as rainfall \citep{Low1982, Barros2008, Villermaux2010}, sea spray \citep{Deike2018, Shaw2024} and volcanic plumes \citep{Jones2019}, and it also plays a critical role across a broad spectrum of engineering applications, underpinning technologies such as combustion systems \citep{lefebvre2017atomization, Wu2023}, nuclear reactor cooling \citep{foissac2011droplet}, spray drying \citep{ameri2006spray}, and ink-jet printing \citep{vanderBos2014, Planchette2019}. In engineering settings, atomization is typically achieved by forcing a liquid through a nozzle, producing a spray of droplets. The atomization process of such an isolated single-nozzle spray has been extensively investigated, focusing on the breakup of the liquid jet \citep{Delon2018, Speirs2020} or liquid sheet \citep{rizk1985spray,Sivakumar1996,Kim2007,Jia2022}, and the subsequent formation of ligaments and droplets \citep{marmottant2004spray, Jalaal2012, Planchette2019, Thivenaz2022}. \textcolor{black}{While much attention has been given to understanding single spray ($SS$), where droplets generally do not experience significant crisscross interactions, many natural and industrial scenarios involve intersecting droplet paths.} In nature, such interactions can occur in rainfall or sea spray, while in engineering systems, they are common in multi-nozzle systems. For instance, the RD-107 engine used in Soyuz rockets consists of 337 pressure swirl-type nozzles arranged in 10 rings \citep{sutton2006history}. The sprays formed from individual nozzle elements interact and mix to form a combined spray ($CS$).  \textcolor{black}{Under atmospheric ambient conditions, such interaction enhances air entrainment and alters the size and velocity characteristics of $CS$ \citep{hardalupas1996interaction,brenn1998experimental,
YOSHIMURA2015}.} Due to changes in the dynamics of droplets of $CS$, the heat and mass transfer behavior differs from that of $SS$. This alters the absorption and reaction characteristics in the combustion chamber, which in turn alters the composition of formed gases \citep{brenn1998experimental}. Thus, for a detailed analysis of the multi-nozzle combustion, it is essential to understand the dynamics and characteristics of spray droplets in $CS$.

\textcolor{black}{Despite their relevance to both engineered and naturally occurring sprays, experimental studies on inter-spray interactions and their influence on droplet dynamics remain limited.} Notably, \citet{hardalupas1996interaction} investigated sprays from three shear coaxial nozzles arranged in a triangular pattern to elucidate the atomisation process in the preburner of the Space Shuttle’s main engine. The study unveils a 25\% reduction in the Sauter mean diameter (SMD), attributed to the transport of smaller droplets from the surrounding individual sprays towards the axis of the neighboring nozzle. This process reduces the mean droplet diameter by decreasing the relative number of large droplets. A 50\% reduction in flow rate induces the promotion of droplet coalescence downstream of the spray axis, leading to an increase in SMD by 10\%. In the study of binary interaction of straight and inclined pressure hollow cone sprays, \citet{brenn1998experimental} observed that the arithmetic mean diameter increases downstream in the case of $CS$ compared to $SS$, which points towards the increased detection of larger droplets and reduction in smaller droplets, indicating droplet coalescence. Furthermore, an increase in the mean axial velocity of smaller droplets is observed for $CS$, with a more pronounced effect in the inclined spray configuration compared to the parallel one, which is reported to be due to the airflow generated by the interaction of the spray. In the context of improving combustion efficiency and reducing fuel consumption by gasoline nozzles, \citet{YOSHIMURA2015} studied the interaction of three pressure swirl nozzles arranged in a triangular configuration. At low injection pressure, three conical liquid sheets interact and form large droplets, increasing SMD. Conversely, as injection pressure increased, the droplets resulting from the conical sheets began to collide. This interaction resulted in a decrease in the difference in SMD between $SS$ and $CS$ with the increase in injection pressure. SMD was slightly lower for the $CS$ at the highest reported injection pressure. 

 Achieving fine atomization is critical across a wide range of spray applications to maximize the surface area of the droplet ensemble and enhance heat, mass, and momentum transfer. Larger droplets formed through the primary atomization process induced by nozzle geometry and flow conditions undergo breakup in the ambient environment, referred to as secondary atomization, and produce numerous smaller droplets. \textcolor{black}{However, fine atomization is affected by droplet collisions in multi-nozzle systems, as the preceding experimental studies highlight the profound impact of such collisions on size and velocity characteristics. Similar effects can also occur in natural sprays. This emphasizes the necessity for a comprehensive understanding of the droplet collision characteristics of $CS$ in both industrial and natural contexts.} While collision among numerous droplets of different sizes is likely to occur in $CS$, the numerical studies have provided reasonable predictions by considering binary collision outcomes \citep{ko2005droplet,kim2009modeling}. 
The outcomes of the binary droplet collision are summarized in previous studies as regime maps that plot impact parameter, $B$, against collision Weber number, $W\!e$, for different size ratios ($\Delta$), which are defined as follows,

\begin{nolinenumbers}
\begin{equation}
 We = \frac{\rho d_s |\vec{v_r}|^2}{\sigma}
\end{equation}
\begin{equation}
\Delta = \frac{d_s}{d_l}
\end{equation}
\begin{equation}
B = \frac{2X}{d_s + d_l}
\end{equation}
\end{nolinenumbers}

\begin{nolinenumbers}
\begin{figure}[!ht]
	\begin{center}		   
    \includegraphics[width=0.55\textwidth,keepaspectratio]{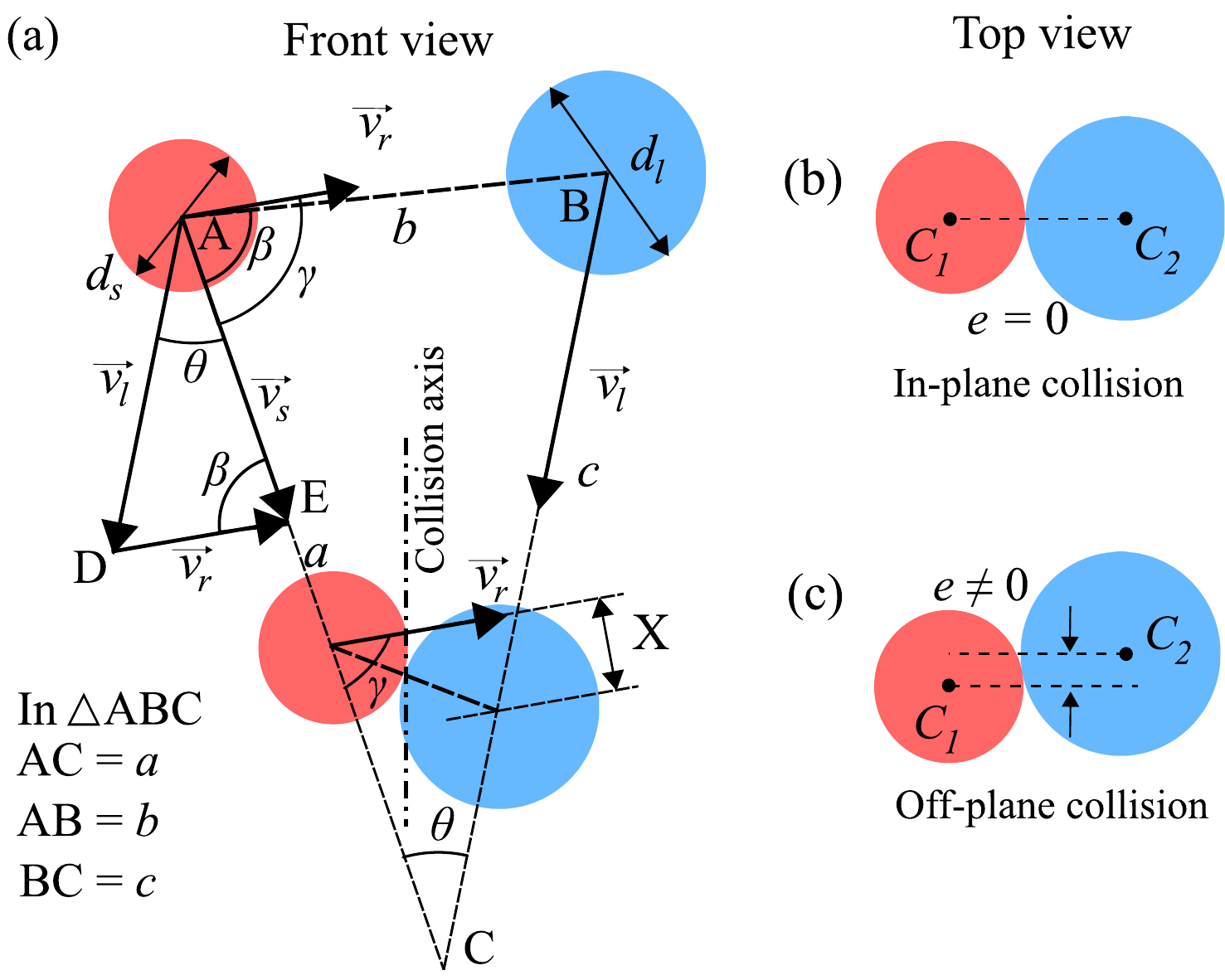}
	\end{center}
   \caption{(a) Schematic illustrating geometrical and kinematic parameters of binary droplet collision. Top view at the point of contact for (b) in-plane collision with eccentricity $e = 0$, and (c) off-plane collision with  $e \neq 0$.}
\label{fig:schematic_coll}
\end{figure}
\end{nolinenumbers}

\hspace{-0.65cm}here $\rho$ and $\sigma$ are the density and surface tension of water, $d_s$ and $d_l$ are the diameters of the smaller and larger colliding droplets, respectively, as shown in Figure \ref{fig:schematic_coll}(a). The magnitude of the relative velocity vector $\vec{v_r}$ between the droplets is calculated using the cosine law in triangle $AED$ as,
\begin{equation}
|\vec{v}_r| = \sqrt{|\vec{v}_s|^2 + |\vec{v}_l|^2 - 2 |\vec{v}_s| |\vec{v}_l| \cos(\theta)}
\end{equation}
here  $v_s$ and $v_l$ are the velocities of smaller and larger droplets, respectively, and $\theta$ is the included angle made by droplet trajectories. $B$ defines the eccentricity of droplets for an in-plane collision, i.e., when both droplet centers are in the same plane concerning the observer, as shown in the top-view of the collision at the point of contact in Figure \ref{fig:schematic_coll}(b). In contrast, centers are in different planes for the off-plane collision as highlighted by eccentricity, $ e \neq 0$ in Figure \ref{fig:schematic_coll}(c). $X$ is the distance between the center of one droplet and  $\vec{v_r}$, which is placed at the center of the other droplet when at contact. Thus, $B$ is calculated as,
\begin{equation}
B = \frac{2b \sin (|\beta - \gamma|)}{d_s + d_l}
\end{equation}
here $\beta$ is the angle between sides $a$ and $b$ and is calculated using the cosine law in triangle $ ABC$ as,
\begin{equation}
\beta = \cos^{-1} \left( \frac{a^2 + b^2 - c^2}{2ab} \right)
\end{equation}
and $\gamma$ is the angle between sides $a$ and $\vec{v}_r$, calculated using the sine law in triangle $AED$ as
\begin{equation}
\gamma = \sin^{-1} \left( \frac{|\vec{v}_l|}{|\vec{v}_r|}  sin \theta\right).
\end{equation}
Although droplet collision outcomes are typically characterized using $W\!e$ with the smaller droplet diameter as the characteristic length scale, \citet{rabe2010experimental} introduced a modified form, known as the symmetric Weber number ($W\!e_s$), to account for the influence of the larger droplet in unequal-size binary collisions. $W\!e_s$ is simply the ratio of total inertial to surface energies of two droplets and can be rewritten as the product of the conventional $W\!e$ and the function of $\Delta$, as,
\begin{equation}
 We_s= f(\Delta) We, \quad f(\Delta) = \frac{\Delta^2}{12 (1 + \Delta^3)(1 + \Delta^2) } 
\end{equation}

The outcomes of binary droplet collision are broadly identified as \textit{bouncing}, \textit{coalescence}, and \textit{reflexive and}  \textit{ stretching separation} in experiments conducted with carefully arranged droplet generators and visualization apparatus \citep{ashgriz1990coalescence,jiang1992experimental,qian1997regimes,orme1997experiments,rabe2010experimental, Jia2019, al2021inertial,  Sui2023}, and numerical studies \citep{pan2005numerical,munnannur2007new,pan2008experimental, Chowdhary2020} for  $W\!e$ of $\mathcal{O}$($10^2$). Bouncing occurs when the collisional kinetic energy is insufficient to expel the gas entrapped between the colliding droplets \citep{orme1997experiments}. \textcolor{black}{Bouncing is not observed for water droplets at atmospheric pressure but is reported at elevated pressures \citep{qian1997regimes}}. When $W\!e$ is large enough, it causes thinning of the air between the droplets to a critical value, eventually leading to their coalescence into a larger droplet. The coalesced droplet may be stable 
or unstable depending upon the values of $W\!e$ and $B$. The separation occurs when the temporarily coalesced droplet breaks apart into satellite droplets. Reflexive separation occurs at nearly head-on collision, i.e., lower $B$, while stretching separation occurs at higher $B$ \citep{ashgriz1990coalescence}. It is worth mentioning that several experimental \citep{santolaya2010analysis,saha2012breakup,santolaya2013effects,wu2021droplet,jedelsky2024effects} and numerical \citep{post2002modeling,ruger2000euler,sommerfeld2019advances, Finotello2019} studies have identified these regimes in $SS$.  

The collision outcomes discussed above primarily pertain to relatively $W\!e$. However, it is important to note that in practical spray systems such as combustion chambers in liquid rocket engines, liquid bulk Weber numbers can reach magnitudes on the order of $\mathcal{O}(10^5)$ \citep{anderson1995impinging}, implying that high-energy droplet collisions with $W\!e$ exceeding $\mathcal{O}(10^2)$ are likely to occur. At higher $W\!e$, head-on binary droplet collision leads to the expansion of a lamella enclosed by a Taylor-Culick (TC) rim. Rayleigh–Taylor instability initiates node and finger formation on the TC rim, which subsequently undergoes capillary (Rayleigh–Plateau) breakup into droplets \citep{Kulkarni2023} at high enough $W\!e$.  This type of binary droplet collision is called \textit{spatter} \citep{Gunn1965}, \textit{splashing} \citep{roth1999high,roth2007droplet}, or \textit{splattering} \citep{pan2009binary}. \citet{roth2007droplet} reported a high energy collision outcome referred to as \textit{stretching with digitations}, which is observed at the mid-range of $B$. Due to the combined effect of inertia and eccentricity, an elliptical rim with thicker ends is ejected in the direction of the relative velocity vector. The rim stretches out, collapses into a ligament, and forms multiple satellite droplets. In the regime map of binary droplet high energy collision of $\mathcal{O}$($10^3$) \citep{roth2007droplet,Zhou2022}, the collision outcomes are identified as splashing, stretching with digitation, reflexive and stretching separation. The splashing regime occurs at a lower $B$ but can be observed at higher $B$ with the increase in $W\!e$. For very high $B$, the stretching separation is dominantly observed for the entire range of $W\!e$.

Although the outcomes of droplet collision are well studied with carefully arranged droplet generators and visualization apparatus, the current literature lacks a similar analysis of the outcomes of colliding droplets under complex poly-disperse spray conditions, particularly for high-energy droplet collisions. Notably, the recent work by \citet{Ghosh2025} has provided valuable insights into collision dynamics in the case of binary interacting sprays from gas-centered coaxial atomizers (\textit{GCSC}), revealing phenomena such as stretching separation and enhanced axial velocity. We employ hollow cone sprays generated from pressure-swirl nozzles to isolate the intrinsic features of droplet collision phenomena from the secondary airflow effects, such as those introduced by a central air jet. These nozzles provide a simplified and well-characterized flow environment that facilitates a fundamental understanding of high-energy droplet collisions in poly-disperse sprays. By eliminating the influence of external airflow, the observed collision outcomes can be more directly compared with binary droplet collision results available in the literature. The three identical nozzles are arranged in a triangular configuration, a simplified and repeatable pattern inspired by nozzle arrangements like those in liquid rocket engines, yet chosen to facilitate fundamental exploration of droplet collision phenomena relevant to atomization processes in both natural and industrial settings. We focus specifically on the most interacting zone, where droplet collisions are prevalent. By utilizing Phase Doppler Interferometry (PDI) and microscopic high-speed backlight imaging, we provide a detailed understanding of droplet collisions in poly-disperse sprays, including the size and velocity characteristics of satellite droplets formed during these events.  The knowledge of droplet collision outcomes and satellite droplet characteristics seen in $CS$ is beneficial in selecting suitable droplet collision models for the numerical simulation of interacting sprays and helps to make improved predictions of droplet characteristics \citep{sommerfeld2019advances}.

\section{Experimental setup and methodology}

\hspace{0.65cm} Figure \ref{fig: schematic}(a) shows the self-explanatory exploded view of the nozzle $CAD$ model. The diameter of the nozzle exit, $d_0$, is 0.57 mm, measured using the optical microscope. The three identical pressure-swirl nozzles (\textit{Spraytech Systems, India}) are labeled as $N1$, $N2$, and $N3$. The nozzles are fastened to elliptical adapters, which are then positioned along three main slots and secured using bolts through side slots provided in the nozzle mounting plate, as shown in Figure \ref{fig: schematic}(b). These slots allow the nozzles to be slid and re-positioned, enabling variation in the spacing, $G$, to achieve an equilateral triangular configuration. In the current work, $G = 30$ mm is maintained to ensure the liquid films from the individual nozzles do not collide. Instead, only the spray droplets interact after a certain vertical distance, $z$, from the nozzle exit plane.

\begin{nolinenumbers}
\begin{figure}[!ht]
	\begin{center}		   
    \includegraphics[width=\textwidth,keepaspectratio=true]{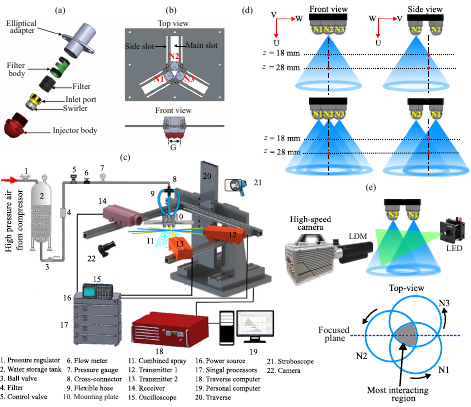}
	\end{center}
   \caption{ (a) Exploded view of the $CAD$ model of the hollow cone pressure swirl nozzle. (b) Triangular arrangement of three nozzles in the mounting plate. (c) Schematic of the spray test facility with phase Doppler interferometer (PDI) and backlight imaging. (d) Locations of PDI measurements in $SS$ (top row) and $CS$ (bottom row) seen from the two views. The dash-dotted line represents the centroidal axis of the arrangement. (e) Schematic illustrating the apparatus used for high-speed microscopic spray visualization. The top view of the spray width diagram shows that the camera is focused on the most interacting region in the U-V plane. The arrows show the anti-clockwise swirl motion of the sprays. }
    \label{fig: schematic}
\end{figure}
\end{nolinenumbers}
\quad \textcolor{black}{The experiments are conducted in an open laboratory environment at atmospheric pressure and room temperature ($\approx$ 298~K). } 
The schematic of the spray test facility is shown in Figure \ref{fig: schematic}(c). The experimental setup consists of an air-compressed liquid storage tank (identified as 2) maintained by a pressure regulator (1), which supplies high-pressure liquid through a filter (4), flow meter (5), control valve (6), and pressure gauge (7) to the nozzle assembly. A push-fit cross-connector (8) links the flexible hoses (9) to the nozzles mounted on the plate (10). The experiments are conducted with water, and the properties (density, $\rho$ = 998 kg/$m^3$, and surface tension, $\sigma$ = 0.0728 N/m at 20$^{\circ}$ C) are taken from \citet{cohen2004fluid}. A 3D Phase Doppler Interferometry (PDI) system ($Artium Technologie$s, USA), a single-point optical diagnostic instrument, is used to simultaneously measure the time-resolved droplet size and all three velocity components (\textit{U},\textit{V}, and \textit{W}). The system is comprised of two transmitters (12 and 13), one receiver (14), an advanced signal analyzer unit (17), a computer-controlled (18) three-axis traverse unit (20), and a computer (19). The focal length of the laser transmitters and receiver lenses is kept at 750 mm and 530 mm, respectively. The transmitters and receiver are positioned on the traverse system so that the receiver is 30 off-axis with respect to the transmitters in forward scatter mode. The present PDI is equipped with AIMS software, automatically choosing optimized receiver settings after examining spray droplets in the measurement volume. The samples are collected for 45 seconds, sufficient to have a data rate of $\mathcal{O}(10^3)$ Hz. Channel 1 validation (validation for the measurements of droplet size and axial velocity) ranges between 70-95 \%, which is above the critical validation threshold of 60\% followed in the previous studies \citep{Tratnig2010}. The measurements are taken along the centroidal axis of the spray arrangement (dash-dotted line), focusing on the most interacting region of $CS$, with a step size of 2 mm over the range $z = 18$–28 mm, as shown in Figure \ref{fig: schematic}(d) (bottom row). Similar measurements are performed for $SS$ from the $N2$ nozzle for comparative analysis, shown in the top row. As observed from the side view of $SS$, the measurement path begins at the periphery of the spray and extends inward, terminating within the hollow core region. The positive directions of the velocity components corresponding to the PDI axes are also indicated in Figure \ref{fig: schematic}(d).

Spray visualization is carried out using the backlighting technique. The system includes a DSLR camera (\textit{Nikon~D7100}) (22) fitted with an A.F. Zoom \textit{Nikkor} 80–200 mm f/2.8D lens, a stroboscope (\textit{Sugawara Laboratories Inc.}, Japan), and a diffuser sheet (21) to produce diffused light pulses at 15 $\mu$s intervals. The pixel array of the camera is 6000 $\times$ 4000, and the stroboscopic images are captured with a resolution of 31.3 $\mu m$ per pixel. The microscopic high-speed videos are recorded using a \textit{Photron SA5} high-speed camera with a long-distance microscope (LDM) (\textit{Questar Corp., USA}) and illuminated by a high-power LED (\textit{Mightex Systems}), as shown in Figure \ref{fig: schematic}(e), to capture collision events in the interacting zone. The camera is focused on the interacting region in the \textit{U-V} plane of PDI as shown by the top-view of the spray width diagram in Figure \ref{fig: schematic}(e). The spray images are captured at 75,000 and 100,000 fps, with pixel arrays of 320 $\times$ 264 and 320 $\times$ 192 pixels, respectively, yielding a $11.7 \mu m$ per pixel resolution. Using the Rayleigh criterion ($DOF =\frac{0.5 \lambda}{\text{NA}^2}$), the depth of field (\textit{DOF}) of LDM calculated with the numerical aperture (NA) of 0.06 and LED wavelength ($\lambda$) of 525 nm is 73 $\mu m$. The exposure time was set at 1$\mu$s to minimize motion blur. ImageJ \citep{Schindelin2012} software is used to measure the size of colliding droplets and track them using the \textit{Manual Tracking plugin}. \textcolor{black}{Considering a one-pixel uncertainty in manual edge selection, the corresponding relative errors range from 6 to 59 \% for droplet sizes between 20 and 200 $\mu m$. The largest relative error corresponds to the smallest droplets. The uncertainty in velocity measurement arises from the spatial resolution of the imaging system and the temporal resolution of 10–13.3~$\mu s$ between successive frames. For an estimated ±1~pixel displacement error, the corresponding uncertainty in velocity is approximately ±(0.9–1.1)~m/s, yielding a relative error of 2–6\% for the measured velocity range of 20–40~m/s. Since the Weber number depends quadratically on velocity, the resulting uncertainty in $W\!e$ is dominated by that of velocity, giving an overall relative error of approximately 12–26\% for droplets ranging from 70 to 200~$\mu m$, for which the Weber numbers are reported.} The accurate classification of binary collisions is inherently challenging due to the stochastic nature of droplet interactions and the trade-off between spatial and temporal resolution in high-speed imaging. In our experiments, $W\!e$ is of $\mathcal{O}(10^{3})$, making it challenging to capture intermediate collision stages owing to camera limitations. Nevertheless, by detecting the temporarily coalesced liquid structure downstream, sufficient information from multiple similar events has been used to identify and characterize the collision regime.
A further challenge is that droplets may collide off-plane (Fig.~\ref{fig: schematic}(c)), i.e., the droplet central plane may not align with the focal plane of the imaging system. \textcolor{black}{ The off-plane collisions cause certain events to appear nearly head-on in the camera plane when they are, in fact, slightly off-center in other orthogonal planes, resulting in off-center collisions. This leads to inaccurate estimations of $B$.} As a result, constructing a regime map from the current data-set is difficult to achieve with high confidence and, therefore, lies beyond the scope of this study.

The nozzles are operated at two injection pressures, $\Delta$P = 550 and 830 kPa, and the respective mass flow rate, $\dot{m}$, is measured using the flow meter. Weber number of the liquid sheet, $W\!e_l$, at the nozzle exit is calculated as, 
\begin{equation}
We_l = \frac{\rho u_l^2 t_f}{\sigma}
\end{equation}
here $u_l$ and $t_f$  are streamwise velocity (along the conical surface) and liquid sheet thickness at the nozzle exit, respectively. $u_l$ is calculated using the cone angle of the spray, $\alpha$, as 
\begin{equation}
u_l = \frac{u_a}{\cos\left(\frac{\alpha}{2}\right)}
\end{equation}
here $u_a$ is the axial velocity at the nozzle exit, calculated from the conservation of mass \citep{Kulkarni2010} as
\begin{equation}
u_a = \frac{\dot{m}}{\rho \pi t_f (d_o - h_f)}.
\end{equation}
$\alpha$ is determined from macroscopic images of the spray. The liquid sheet thickness ($h_f$) is calculated from the nozzle geometry and flow parameters as proposed by \citet{Suyari1986}, given by,
\begin{equation}
t_f = 2.7 \left( \frac{\dot{m} d_o \mu}{\rho \Delta P} \right)^{0.25}.
\end{equation}
The values of spray flow parameters are presented in Table \ref{tab:experimental_data}.

\begin{nolinenumbers}
\begin{table}[h]
    \centering
    \renewcommand{\arraystretch}{1.2}
    \begin{tabular}{cccccccc}
        \toprule
        S. No & $\Delta P$ (kPa) & $\dot{m}$ (g/s) & $\alpha$ (°) & $u_a$ (m/s) & $u_l$ (m/s) & $t_f$ ($\mu$m) & $W\!e_l$ \\
        \midrule
        1 & 550  & 4.05 ± 0.05  & 89.7 ± 2.5 & 23.3  & 33.3 & 122.6  & 1896 \\
        2 & 830 & 4.97 ± 0.12  & 84.7 ± 2    & 30.3  & 41.2  & 115.8  & 2704 \\
        \bottomrule
    \end{tabular}
    \caption{Flow parameters for two test conditions.}
    \label{tab:experimental_data}
\end{table}
\end{nolinenumbers}

\section{Results and discussion}

 \hspace{0.65cm}The stroboscopic images of $CS$ from the multi-nozzle arrangement at $We_l = 2704$ are shown in Figure \ref{fig:spray_strobo}, from two perspectives: front view (Fig. \ref{fig:spray_strobo}(a)) and side view (Fig. \ref{fig:spray_strobo}(b)). Droplets formed from the breakup of conical liquid sheet from individual nozzles begin to mix around $z/G = 0.6$, resulting in a dense droplet cloud along the centroidal axis (dash-dotted line). The position of the centroidal axis relative to nozzles $N2$ and $N3$ is visible in the side view (Fig.~\ref{fig:spray_strobo} (b)), clearly indicating the location of the dense droplet cloud formed by the mixing of all three sprays. This spatial reference helps identify the most interacting zone. Additionally, the shaded region in the spray width diagram (Fig.~\ref{fig: schematic} (e)) highlights the top view of the highly interacting region.

\begin{nolinenumbers}
\begin{figure}[!ht]
	\begin{center}		   
    \includegraphics[width=\textwidth,keepaspectratio=true]{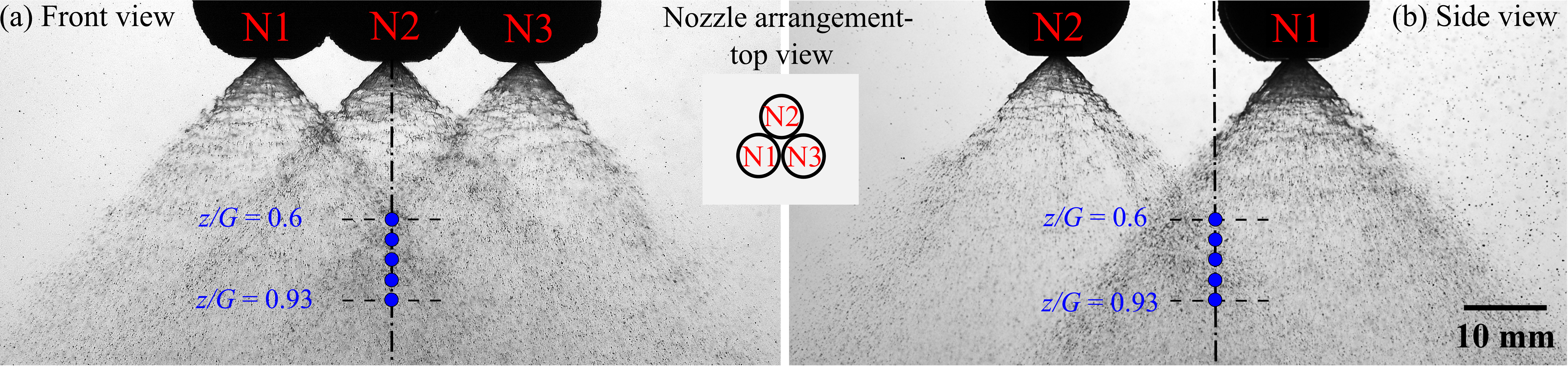}
	\end{center}
   \caption{ Stroboscopic images of $CS$ from the multi-nozzle captured from (a) front, and (b) side views at $W\!e_l$ = 2704. In the side view, the spray from the nozzle element $N3$ appears behind that from the nozzle $N1$, as can be seen in the top-view of the nozzle arrangement. The dash-dotted line is the centroidal axis.}
    \label{fig:spray_strobo}
\end{figure}
\end{nolinenumbers}

\begin{nolinenumbers}
\begin{figure}[!h]
	\begin{center}		   
    \includegraphics[width=0.85\textwidth,keepaspectratio=true]{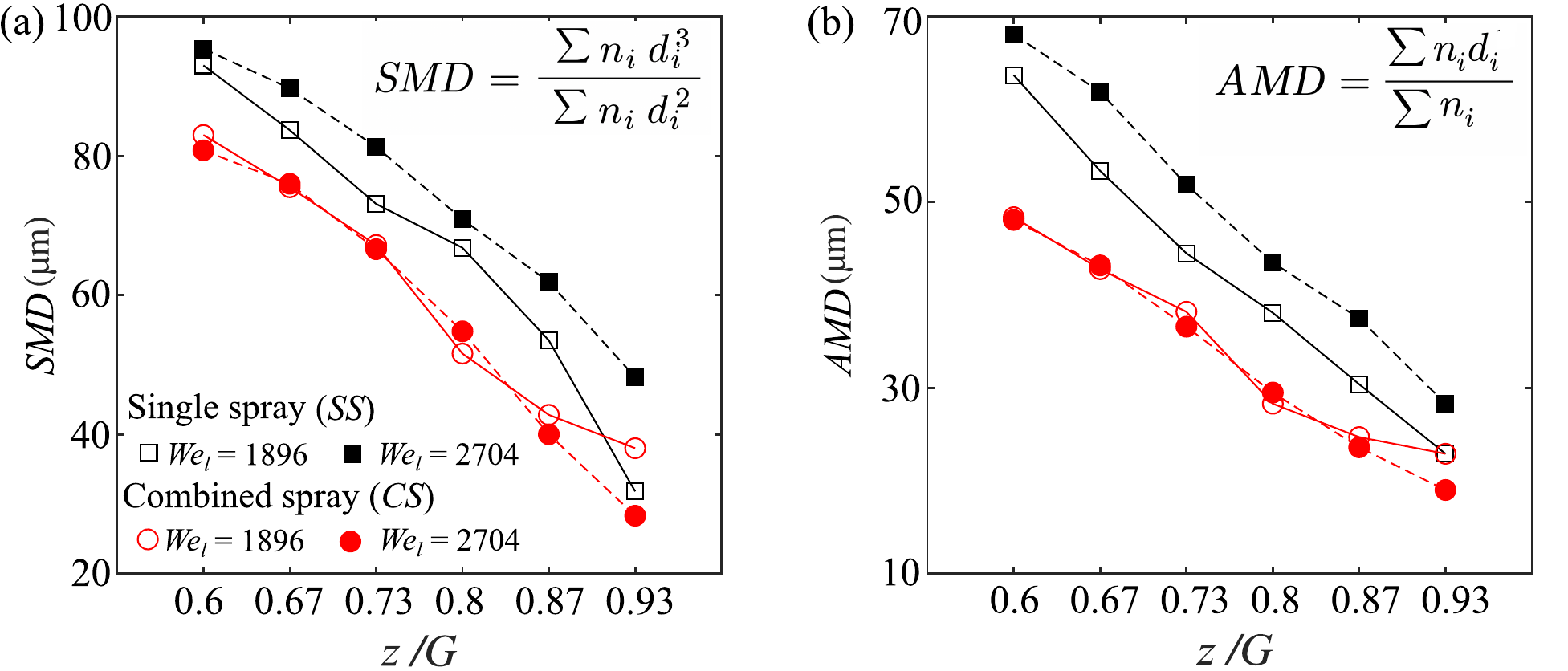}
	\end{center}
    \vspace{-10pt} 
   \caption{(a) SMD and (b) AMD variation at different $z/G$ along the centroidal axis for $SS$ and $CS$ at different $W\!e_l$. The expressions of SMD and AMD are mentioned in the respective graphs, here $n_i$ is the number of droplets with diameter $d_i$. }
   \label{fig:smd_amd}
\end{figure}
\end{nolinenumbers}

To assess the interaction phenomenon in \textit{CS} and compare it with \textit{SS} (from the \textit{N2} nozzle), spray droplet measurements are taken along the centroidal axis (dot-dashed line), as indicated by the points in Figures~\ref{fig: schematic}(d) and~\ref{fig:spray_strobo}. Figure~\ref{fig:smd_amd}(a) presents the Sauter mean diameter (SMD)  profile along the centroidal axis. The SMD decreases nearly linearly with \textit{z/G} in both \textit{SS}  and \textit{CS}. In the case of \textit{SS}, the probe volume traverses from the spray boundary into the less densely populated interior of the hollow-cone spray, as shown in the side view in Figure~\ref{fig: schematic}(d), resulting in a decrease in SMD along the centroidal axis. In \textit{CS}, a similar trend is observed, although the SMD is lower than in \textit{SS} by approximately 11\% at $\mathit{We_l} = 1896$ and 15\% at  $\mathit{We_l} = 2704$ near the onset of interaction.
It is important to note that SMD is highly sensitive to larger droplets, and even a few such droplets within the probe volume can lead to a marked increase in SMD. To better capture the contribution of smaller droplets, the arithmetic mean diameter (AMD) is also shown in Figure~\ref{fig:smd_amd}(b), as it more clearly reflects number-based changes. A reduction in AMD of 24\% at $\mathit{We_l} = 1896$ and 30\% at $\mathit{We_l} = 2704$ is observed near the interaction onset, highlighting the increased presence of smaller droplets due to interaction dynamics.

\begin{figure}[!h]
	\begin{center}		   
    \includegraphics[width=0.75\textwidth,keepaspectratio=true]{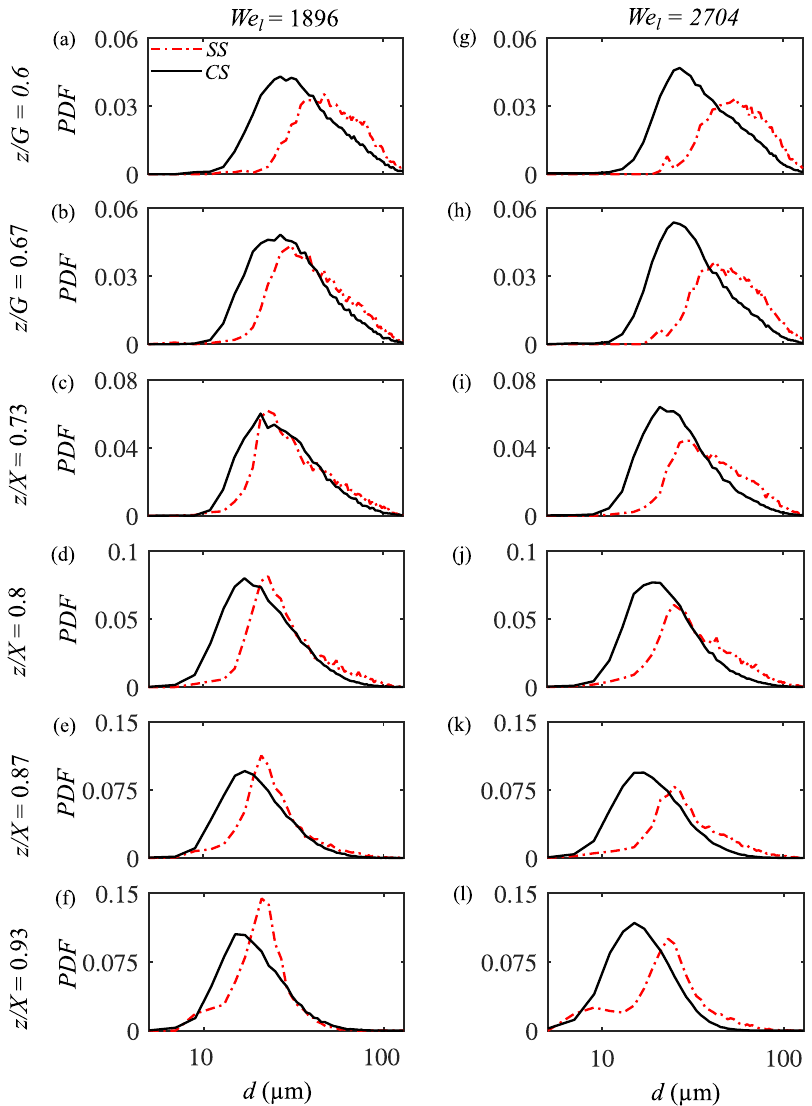}
	\end{center}
    \vspace{-10pt} 
   \caption{Droplet-size distribution comparison between \textit{SS} and \textit{ CS} at different \textit{z/G} for $W\!e_l$ = 1896 (a-f) and $W\!e_l$ = 2704 (g-l).  }
   \label{fig:droplet_dis}
\end{figure}

The SMD and AMD trends can be better understood from the droplet size distribution shown in Figure \ref{fig:droplet_dis}. The PDF is calculated by normalizing the number count for each size class by taking a bin size of 2 $\mu m$. \textcolor{black}{Each PDF is constructed from the validated droplets recorded in Channel~1, totaling approximately $5\times10^3$ to $1.2\times10^5$ droplets, which is sufficient to achieve statistical convergence.} Both \textit{SS} and \textit{CS} exhibit uni-modal PDFs, consistent with prior observations \citep{Tratnig2010}. However, a bimodal distribution appears in the hollow region of \textit{SS} (Fig.\ref{fig:droplet_dis} (f)), becoming more pronounced at higher $\mathit{We_l}$ (Fig.     \ref{fig:droplet_dis} (l)), while no such feature is observed in \textit{CS}. Compared to \textit{SS}, \textit{CS} shows a notable increase in the count of smaller droplets and a reduction in larger ones, leading to lower SMD and AMD values. For example, at $\mathit{We_l} = 1896$ and $z/G = 0.6$ (Fig. \ref{fig:droplet_dis}(a)), the number of smaller droplets (25–35 $\mu m$) in \textit{CS} is roughly three times higher than in \textit{SS}, resulting in a local reduction of 11\% in SMD and 24\% in AMD. 
In addition to the previously noted effect of spray divergence in the case of \textit{SS}, droplet collisions become a potentially significant factor in $CS$ as interactions between opposing sprays occur at high $W\!e$ of $\mathcal{O}(10^2$–$10^3)$ (based on calculation using PDI measurements and spray cone angle), sufficient to cause fragmentation. The noticeable reduction in larger droplets supports this hypothesis, indicating that droplets may undergo collisions that generate smaller satellite droplets. Consequently, \textit{CS} warrants detailed microscopic investigation using LDM and high-speed imaging, as illustrated in Figure~\ref{fig: schematic}(e), with a focused examination of the plane intersecting the most interacting region.

    \begin{figure}[!ht]
	   \begin{center}		   
        \includegraphics[width=0.95\textwidth,keepaspectratio=true]{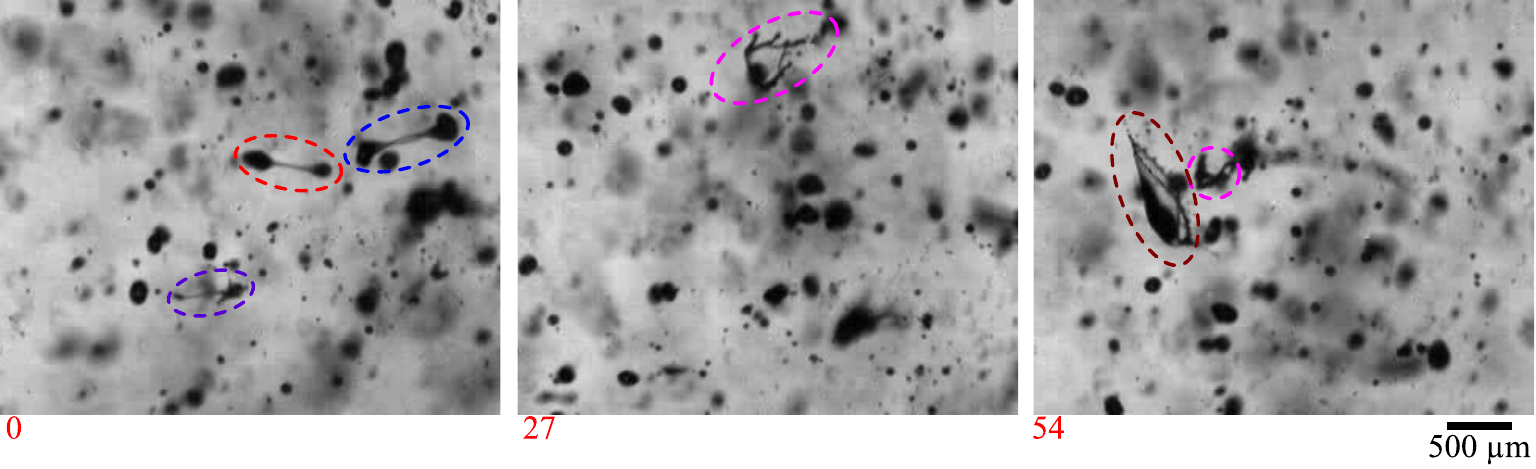}
	   \end{center}
       \caption{Image sequence showing multiple transient coalesced structures from collision events in $CS$ at $W\!e_l$ = 2704. The color of the ellipse shows such transient coalesced structures. Time in $\mu$s is mentioned on the left.  }
     \label{fig:collision_window}
    \end{figure}

\subsection{Droplet collision phenomena}

\begin{figure}[!ht]
	\begin{center}		   
    \includegraphics[width=0.55\textwidth,keepaspectratio=true]{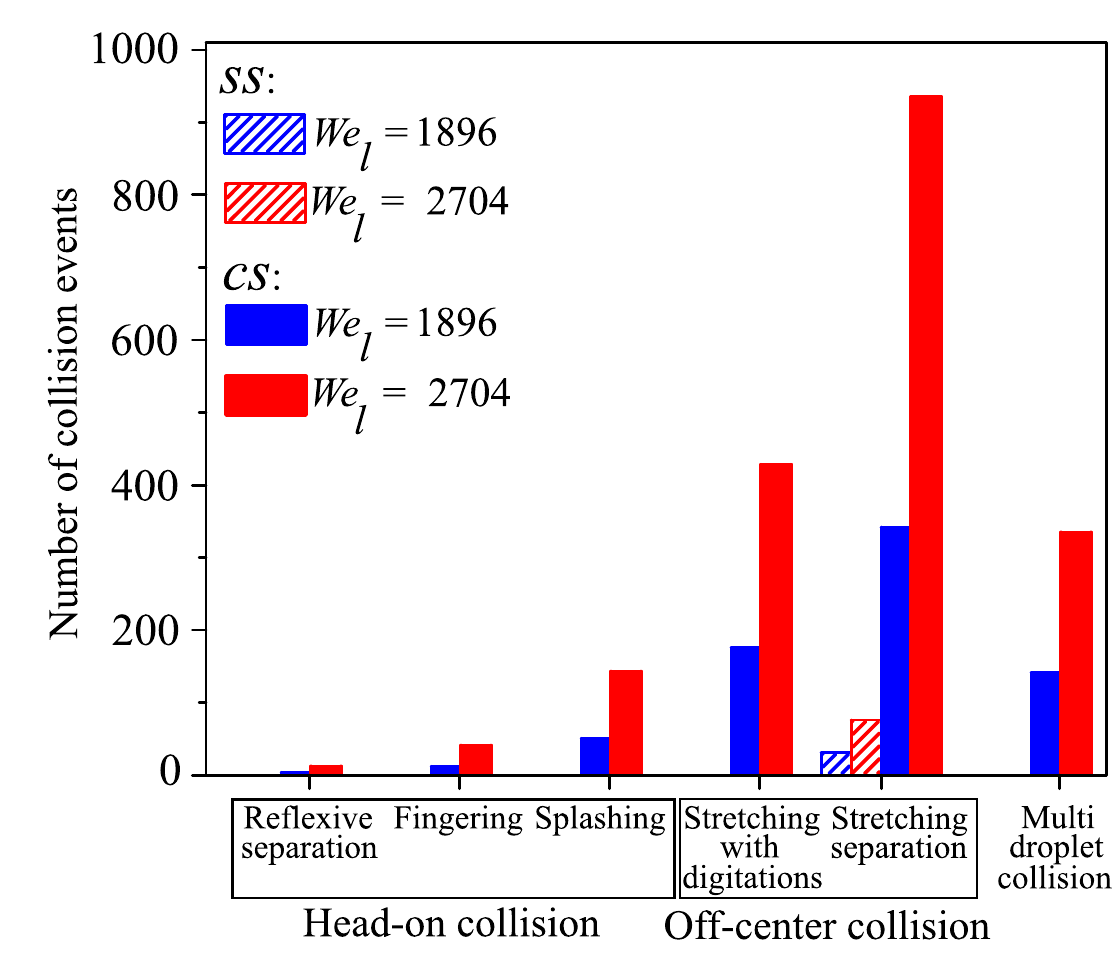}
	\end{center}
    \vspace{-10pt} 
   \caption{Number of droplet collision events in the interacting region of $CS$ and the same region of $SS$ captured within 0.1 sec of recording at 75,000 FPS. }
   \label{fig:collision_freq}
\end{figure}

    The media files \href{https://drive.google.com/file/d/1QixnZAbveNYv6fMaBbELrWVZj_x-l3hF/view?usp=sharing}{$M\!F\!S\!S1$} and \href{https://drive.google.com/file/d/1ioiG44egnzBJ5KpOi9C1T3csN0OmAM87/view?usp=sharing}{$M\!F\!S\!S2$} , and \href{https://drive.google.com/file/d/1ODG_nEm9cqdvDIoLWJcS_Hhl9s8VwKJz/view?usp=sharing}{$M\!F\!C\!S1$} and \href{https://drive.google.com/file/d/1tiMP8BZxJ4RhWWFY076ZeTBD4NPMjIhc/view?usp=sharing}{$M\!F\!C\!S2$}  (see supplementary material), show high-speed videos of $SS$ and $CS$, respectively, captured at 75,000 FPS. These recordings focus on $z/G = 0.6$–0.7, covering the interaction zone in \textit{CS} and the corresponding region in \textit{SS} for comparison. Each pair of videos corresponds to $\mathit{We_l} = 1896$ and $\mathit{We_l} = 2704$, respectively. The collision events captured in the high-speed $CS$ videos, as further illustrated in the image sequence in Figure~\ref{fig:collision_window}, exhibit a striking frequency and simultaneity within a very short time interval. The ellipses in  Figure~\ref{fig:collision_window} highlight the transient coalesced structures that emerge during these ongoing different collision events. Figure~\ref{fig:collision_freq} shows the collision frequency at two $W\!e_l$ values, based on analysis of 7500 frames captured over 0.1 s at 75,000 FPS.  $CS$ exhibits a markedly higher frequency of collisions compared to $SS$. In \textit{SS}, only a few stretching separation events are observed (see media $M\!F\!S\!S1$ and $M\!F\!S\!S2$), whereas \textit{CS} spans the full range of collision outcomes. Although the macroscopic view in Figure~\ref{fig:spray_strobo} suggests that the interaction region is densely packed, high-speed videos reveal that the droplet sizes are much smaller than the inter-droplet spacing. This condition overwhelmingly favors off-center impacts, namely, \textit{stretching with digitations} and \textit{stretching separation} over head-on collisions. Head-on impact results in either \textit{reflexive separation}, \textit{fingering}, and \textit{splashing}, depending on the collision $W\!e$. One or more additional droplets may impact the temporarily coalesced structure, resulting in multi-droplet collision events. Such interactions are also observed and occur more frequently than head-on collisions, principally because the coalesced liquid structure provides a substantially larger surface area, greatly increasing the likelihood of droplet impingement. Increasing $\mathit{We_l}$ raises all collision outcomes, with off-center and multi-droplet collisions far outnumbering head-on collisions. While the current data represent specific test conditions, the observed collision frequencies may vary with changes in droplet density, size, velocity distributions, and the value of $G$.

    The following section discusses the collision dynamics for different collision outcomes since such high-energy collisions are sparsely reported in the literature. A key objective is to quantify the characteristics of satellite droplets formed under various collision scenarios, and to use these findings to highlight the distinctions between \textit{CS} and \textit{SS}. The discussion begins with binary droplet collisions, while multi-droplet interactions are addressed later.

\begin{nolinenumbers}
\begin{figure}[!ht]
	\begin{center}		   
    \includegraphics[width=\textwidth,keepaspectratio=true]{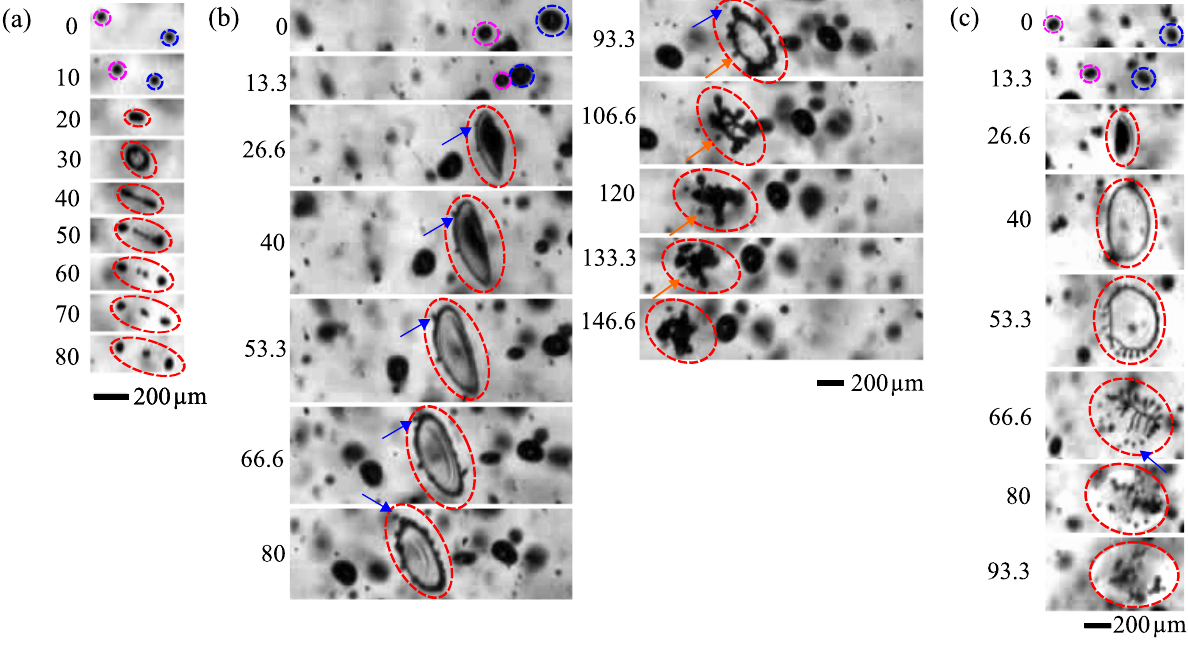}
	\end{center}
    \vspace{-10pt} 
   \caption{Image sequences illustrating head-on binary droplet collisions in $CS$. The cases shown correspond to: (a) \textit{reflexive separation} between droplets at $W\!e_s$ = 4.7 ($W\!e$ = 217, $v_r$ = 20 m/s, $d_L$ = 47 $\mu$m, $d_R$ = 40 $\mu$m, and $W\!e_l$ = 1896), (b)  \textit{Fingering} at $W\!e_s$ = 19.3 ($W\!e$ = 959, $v_r$ = 26.7 m/s, $d_L$ = 98.5 $\mu$m, $d_R$ = 144 $\mu$m and $W\!e_l$ = 1896), (c) \textit{Fingering} at $W\!e_s$ = 28.0 ($W\!e$ = 1306, $v_r$ = 36.9 m/s, $d_L$ = 70 $\mu$m,  $d_R$ = 75 $\mu$m and $W\!e_l$ = 2704). Here, $d_L$ and $d_R$ are the left and right droplet diameters, respectively. The smaller one is used to calculate $W\!e$. Time in $\mu$s is mentioned on the left. The high-speed videos of the events are shown in \href{https://drive.google.com/file/d/1Z-SHutyxM6MgGmbc_O-BwV1seCXh2pxW/view?usp=sharing}{$MF-8a$}, \href{https://drive.google.com/file/d/1iE5x25C0yEnr0bn2wOLFMSapfwASPUD3/view?usp=sharing}{$MF-8b$}, and \href{https://drive.google.com/file/d/1IYkWBLYhraG-_B4DFHm66_5bnXxWYLVK/view?usp=sharing}{$MF-8c$}. In videos, left and right droplets are referred to as parent $P1$ and $P2$, respectively.}
    \label{fig:head_on_hs}
\end{figure}
\end{nolinenumbers}
\vspace{0.25cm}

\subsubsection{\textbf{Binary head-on collision}}  \label{sec:head_on} 
\hspace{-0.15cm}\underline{\textit{Reflexive separation}}
\vspace{0.5cm}

When the droplets collide head-on, i.e., when $B$ is minimal, a torus-like ring bounding a liquid lamella is formed \citep{ashgriz1990coalescence}. The fate of this structure depends on the magnitude of $W\!e_s$. Figure \ref{fig:head_on_hs}(a) shows the image sequence of the binary collision of spray droplets with $W\!e_s$ = 4.4 ($W\!e$ = 217). The colliding droplets squeeze each other, forming a lamella bounded by the TC rim due to a surface tension-driven end-pinching effect ($t = $ 30 $\mu$s) \citep{kuan2014study}. As time progresses, due to insufficient inertial force for further expansion, the rim collapses into a ligament ($t = $ 40 $\mu$s) due to the reflexive action of surface tension. The ligament breaks into three satellite droplets (seen at $t = $ 50 - 80 $\mu$s). This phenomenon is termed \textit{reflexive separation} \citep{orme1997experiments,ashgriz1990coalescence}.

\hspace{-0.55cm}\underline{\textit{Fingering}}

\vspace{0.25cm}
For a similar head-on collision of droplets with higher inertial force,  as shown by the collision at $We_s = 19.3$ in Figure \ref{fig:head_on_hs}(b), the droplets collide and begin to expand to form a lamella bounded by a thick rim. This rim is accelerated radially and in accordance with Rayleigh-Taylor instability \citep{Kulkarni2023} form corrugations called nodes shown by an arrow at $t=$ 26.6 $\mu$s. Surface tension enables the selection of the most destructive wavelength for these corrugations. As the rim expands, the corrugations along its circumference grow radially, resulting in the growth of fingers on the rim as highlighted by the blue arrow from $t=$ 26.6 - 66.6 $\mu$s. As the rim is unable to expand further, the rim and fingers recede, resulting in the collapse of the lamella ($t = $ 120 $\mu$s). The retraction of fingers results in the ejection of satellite droplets, as highlighted by the blue arrows at $t = $ 66.6 – 93.3 $\mu$s and orange arrows at $t = $ 93.3 - 133.3 $\mu$s. The observed collision behavior is termed in the literature as $\textit{fingering}$ \citep{kuan2014study}. With a further increase in inertial forces, as shown in Figure~\ref{fig:head_on_hs}(c) at $We_s = 28$, fingers begin to form rapidly within a short time frame ($t = 53.3\ \mu\text{s}$). Surface tension causes the rim to retract almost immediately ($t = 53.3 \text{--} 66.6\ \mu\text{s}$), leading to the formation of a branched liquid structure with multiple elongated fingers. These fingers subsequently eject satellite droplets, as indicated by the arrows at $t = 66.6\ \mu\text{s}$. The branched ligament continues to disintegrate, producing additional satellite droplets observed at $t = 80$ -- $93.3\ \mu\text{s}$.

\begin{nolinenumbers}
\begin{figure}[!ht]
	\begin{center}		   
    \includegraphics[width=\textwidth,keepaspectratio=true]{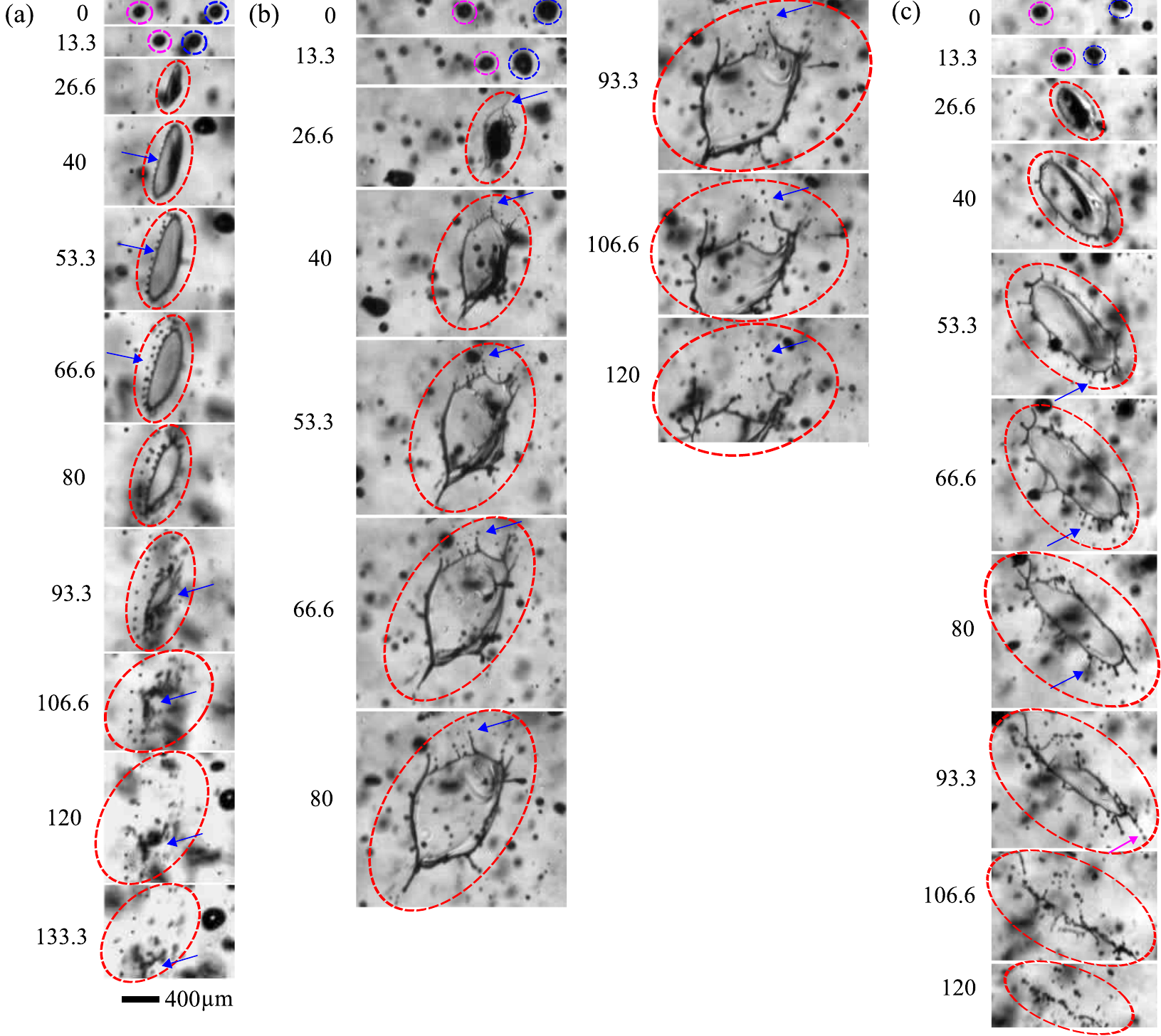}
	\end{center}
    \vspace{-10pt} 
   \caption{{Image sequences illustrating  $\textit{high energy splashing}$ from head-on binary droplet collisions in $CS$. The cases shown correspond to:  (a) $W\!e_s$ = 33.3 ($W\!e$ = 1539, $v_r$ = 32.4 m/s, $d_L$ = 106 $\mu m$, $d_R$ = 126 $\mu m$, and  $W\!e_l$ =1896  ), (b) $W\!e_s$ = 57.4 ($W\!e$ = 2709, $v_r$ = 35.9 m/s, $d_L$ = 151 $\mu m$, and $d_R$ = 199 $\mu m$, and  $W\!e_l$ = 2704), and (c) $W\!e_s$ = 62.8 ($W\!e$ = 2939, $v_r$ = 39.5 m/s, $d_L$ = 136 $\mu m$ and $d_R$ = 144 $\mu m$ , and  $W\!e_l$ = 2704). Time in $\mu$s is mentioned on the left. {The scale bar for all cases is the same.} The high-speed videos of the events are shown in \href{https://drive.google.com/file/d/1QCf4CCBU_xvxJt-wU_r04nvs8mixmN-L/view?usp=sharing}{$MF-9a$}, \href{https://drive.google.com/file/d/1c_cR_BZNFntK1eLp-QsQi7U6iqCIBv6N/view?usp=sharing}{$MF-9b$}, and \href{https://drive.google.com/file/d/12UKT2r8EJmpBnhGAgKza9z9UKoS2r5qa/view?usp=sharing}{$MF-9c$}. } }
    \label{fig:head_on_high}
\end{figure}
\end{nolinenumbers}

\hspace{-0.55cm}\underline{\textit{Splashing}}

 \vspace{0.25cm}
 At higher $W\!e_s$, finger formation and satellite droplet ejection from the rim occur much earlier in the expansion stage of the rim. This is attributed to the insufficient surface energy to contain such high kinetic energy present in the colliding droplets. Such droplet morphological behavior is similar to the splashing phenomenon found in the case of single droplet impact on a solid surface \citep{Xu2005,Opfer2014,Josserand2016}. Figure~\ref{fig:head_on_high}(a) presents the image sequence of a head-on binary droplet collision at $We_s = 33.3$. The lamella rapidly expands, and satellite droplets are ejected from the rim, as indicated by the arrow at $t$ =  66.6 $\mu$s. Subsequently, both the rim and lamella begin to recede ($t$ =  80  $\mu$s), eventually collapsing into a central liquid mass that disintegrates into smaller droplets at later stages, as indicated by the arrows at $t$ =  93.3 -- $120\ \mu$s. Figure \ref{fig:head_on_high}(b) and (c) illustrate the spray droplets collide at much higher inertia, with $W\!e_s$ = 57.4 and 62.8, respectively. It can be seen that the fingers originate on the TC rim instantly after impact and eject multiple satellite droplets from same finger (indicated by arrows at $t = $ 26.6 - 120 $\mu s$ in Figure \ref{fig:head_on_high}(b), and $t$ =  53.3 - 80 $\mu s$ (blue) and $t$ =  93.3  $\mu s$ (magenta) in Figure \ref{fig:head_on_high}(c)). High inertia of the colliding droplets leads to the formation of leaner and longer fingers during the expansion of the rim.
 
\begin{nolinenumbers}
\begin{figure}[!ht]
	\begin{center}		   
    \includegraphics[width=0.8\textwidth,keepaspectratio=true]{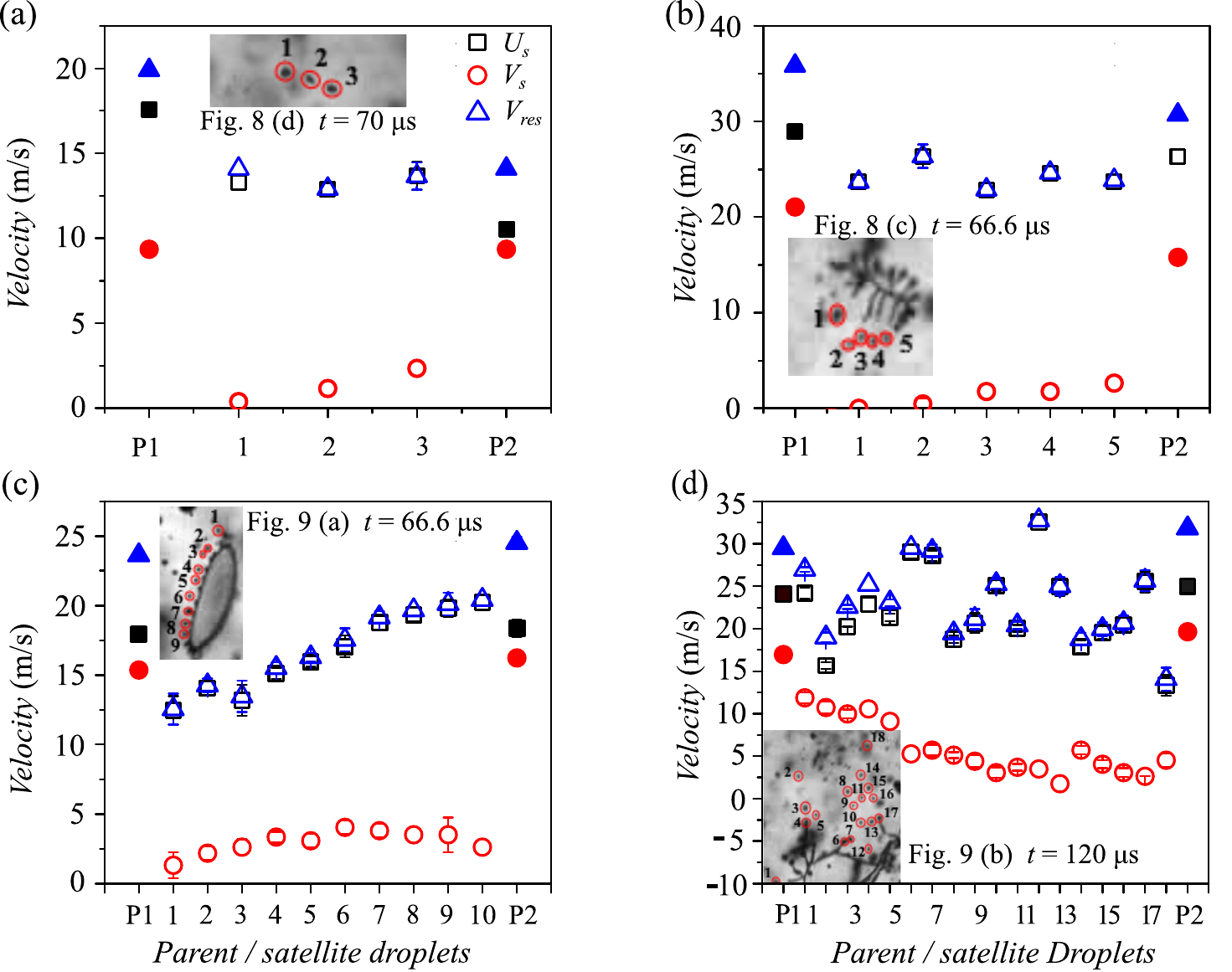}
	\end{center}
    \vspace{-10pt} 
   \caption{Measured axial ($U_s$) and radial velocities ($V_s$) of parent (filled symbols) and satellite droplets (open symbols) for the head-on droplet collisions shown in Figures \ref{fig:head_on_hs} and \ref{fig:head_on_high}. ($a$) \textit{Reflexive separation} shown in Figure  \ref{fig:head_on_hs}(a). ($b$) \textit{Fingering} shown in  Figure \ref{fig:head_on_hs}(c). ($c$)  \textit{Splashing} shown in  Figure \ref{fig:head_on_high}(a). ($d$) \textit{higher energy splashing} shown in  Figure \ref{fig:head_on_high}(b). $V_{res}$ is the resultant velocity. The number denotes the satellite droplets, annotated in the supplementary videos and shown in the graphs themselves. \textit{P1} and \textit{P2} denotes the parent droplets.} 
   \label{fig:vel_splashing}
\end{figure}
\end{nolinenumbers}

 Upon meticulous examination of the image sequences, the axial ($U_s$) and radial ($V_s$) velocities of the satellites are measured. Figure \ref{fig:vel_splashing} presents the $U_s$ and $V_s$ velocities of satellites formed from head-on collisions shown in Figures \ref{fig:head_on_hs} and \ref{fig:head_on_high}. 
 Each graph includes an inset image showing the identity and ordering of the satellite droplets, which is annotated in supplementary medias. The x-axis labels correspond to the numbered satellite droplets in the image. While $P1$ (left) and $P2$ (right) indicate the parent droplets involved in the collision. Error bars shown in the data represent the standard deviation in velocity, measured across multiple frames. However, the calculated uncertainties are so minimal that they fall within the symbol size. It can be seen that the satellite droplets exhibit higher $U_s$ than $V_s$ in all the head-on collisions. Consequently, $U_s$ closely aligns with the resultant velocity, $V_{res}$. It is essential to acknowledge that the depth of the field of the imaging setup is confined to 73 $\mu m$. If the tangential velocity (whether directed in or out of the paper) were comparable to other components, tracking satellite droplets in subsequent frames might have proven challenging. This underscores the crucial observation that the trajectory of satellite droplets is predominantly oriented toward the axial direction.

\subsubsection{\textbf{Binary off-center collision}} \label{sec:off_cent} 
\hspace{-0.15cm}\textit{\underline{Stretching separation}}

 Figure  \ref{fig:stretching}(a) illustrates the image sequence for the collision of droplets with a significant off-center impact at $B$ = 0.9 and $W\!e_s$ = 5.5. In this scenario, only a small portion of droplets come into contact, and the parent droplets persist in following their initial trajectories. This leads to the growth of a smooth ligament ($t = $ 30 – 50 $\mu s$). The ligament stretches out and pinches off from the protuberant ends ($t = $ 60 $\mu  s$). The process of pinch-off results in capillary waves, destabilizing the ligament and fragments into multiple tiny satellites ($t = $ 60 - 90 $\mu s$). The observed collision behavior is termed in the literature as \textit{stretching separation} \citep{ashgriz1990coalescence}. Off-plane stretching separation, depicted in Figure \ref{fig:stretching}(b), provides a comprehensive understanding of the regime by capturing the temporarily coalesced structure from different orientations. The collision between a droplet pair at $W\!e_s$ = 12.8 leads to the formation of a liquid lamella bounded by TC rim (highlighted by arrow at $t = $ 30 $\mu s$) that elongates, collapses into a ligament ($t = $ 40 -- 50 $\mu s$), and then disintegrates into an array of satellites ($t = $ 60 – 90 $\mu s$).

\begin{nolinenumbers}
\begin{figure}[!ht]
	\begin{center}		   
    \includegraphics[width=1\textwidth,keepaspectratio=true]{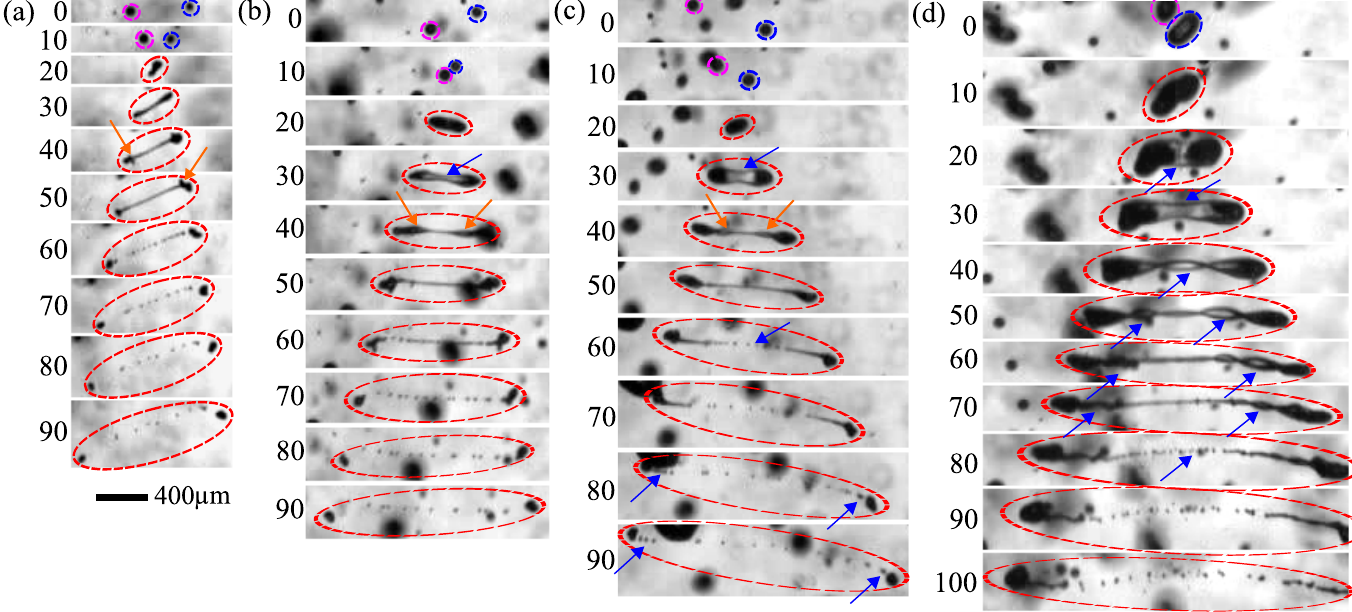}
	\end{center}
    \vspace{-10pt} 
   \caption{{Image sequences illustrating {eccentric binary droplet collisions} in $CS$ at $W\!e_l$ = 1896. The cases shown correspond to {\textit{stretching separation}}:  (a) $W\!e_s$ = 5.5 ($W\!e$ = 305, $v_r$ = 23.2 m/s, $d_L$ = 64 $\mu m$, $d_R$ = 40 $\mu m$, $B$ = 0.9 and $\Delta$ = 0.61), (b) $W\!e_s$ = 12.8 ($W\!e$ = 612, $v_r$ = 26.7 m/s, $d_L$ = 84 $\mu m$, $d_R$ =62 $\mu m$, and $\Delta$ = 0.74), (c) $W\!e_s$ = 29.3 ($W\!e$ = 1359, $v_r$ = 32.4 m/s, $d_L$ = 84 $\mu m$, $d_R$ = 88 $\mu m$, and $\Delta$ = 0.95, and (d) $W\!e_s$ = 49.3 ($W\!e$ = 2283, $v_r$ = 28.9 m/s, $d_L$ =197 $\mu m$, $d_R$ = 220 $\mu m$, and $\Delta$ = 0.89). Time in $\mu$s is mentioned on the left. {The scale bar for all cases is the same.}  The high-speed videos of the events are shown in \href{https://drive.google.com/file/d/1NR7Jc8quVzcNdpbGHL5_A6Fl_vHPdmFI/view?usp=sharing}{$MF-11a$}, \href{https://drive.google.com/file/d/1nKvotTJmzqcATAQk2IyPWd3aTCPFeHT5/view?usp=sharing}{$MF-11b$}, \href{https://drive.google.com/file/d/15lpHzvwYk73NiFMUvBnZe7JrHi3LbqHf/view?usp=sharing}{$MF-11c$}, and \href{https://drive.google.com/file/d/1bFqihuuWBzdlWV3z8GL2HS7Z0NmBZoZW/view?usp=sharing}{$MF-11d$}.} }
    \label{fig:stretching}
\end{figure}
\end{nolinenumbers}
 \vspace{0.25cm}
 
 Figure \ref{fig:stretching}(c) illustrates off-plane stretching separation observed at significantly higher inertia with $W\!e_s$ = 29.3, where the process of lamella formation bounded by TC rim (shown by arrow at $t = $ 30 $\mu s$) and subsequent collapse into ligament ($t = $ 40 - 50 $\mu s$) becomes more evident. \textcolor{black}{Notably, the ligament breakup initiates from the portion (shown by arrow at $t = $ 60 $\mu s$) closer to the collision axis {(indicated as dash-dot line in Fig. \ref{fig: schematic}(a))} rather than the protuberant ends}. In other words, satellites are first formed in the ligament segment near the collision axis, in contrast to the pinch-off from the protuberant ends observed in low $W\!e_s$ cases. The preferential breakup of ligament is likely attributed to the vase shape of the thinner section, characterized by higher curvature variation and, consequently, elevated capillary pressure \citep{Marmottant}. This indicates that strong inertial forces cause the thinning of the ligament section closer to the collision axis. Ultimately, the ligament eventually pinches off from protuberant ends, giving rise to additional satellite droplets, as shown by arrows at $t = $ 80 - 90 $\mu s$.    

Intriguingly, at much higher inertial forces as in Figure \ref{fig:stretching}(d) for $W\!e_s$ = 49.3, the transformation of the TC rim into a ligament is more clearly visible ($t = $ 20 - 30 $\mu s$). Instabilities manifest as knots on the TC rim that increase in number with time (shown by arrows at $t = $ 40 – 70 $\mu s$) before ultimately collapsing into a ligament at $t = $ 70 $\mu s$. Subsequently, the ligament undergoes breakup once again from the region closer to the collision axis (shown by arrow $t = $ 80 $\mu s$), forming multiple smaller satellites. Undoubtedly, the physics governing the transformation of the rim into the ligament is both intriguing and pivotal, as it ultimately dictates the size and velocity distribution of the satellites. The recent work by \citet{Lo2025} on spinning twitted rims formed between two expanding holes on a curved liquid sheet can explain the knots and thus ligament formation. The TC rim, along with the lamella, may be twisted and give the impression of knots. The twisted motion in the case of expanding holes on a curved sheet is due to the lateral collision. In contrast, the twisting motion here may arise because the centers of mass of the protuberant ends do not lie on the longitudinal principal axis of the liquid structure. This offset creates a torque at each end in opposite directions, thereby twisting the rim and facilitating its transition into a filament. It should be noted that in lower $W\!e_s$ cases (Fig.  \ref{fig:stretching}(a--c)), the knots or the twisting of the rim can also be seen as marked by orange arrows.  

\begin{nolinenumbers}
\begin{figure}[ht]
	\begin{center}		   
    \includegraphics[width=1\textwidth,keepaspectratio=true]{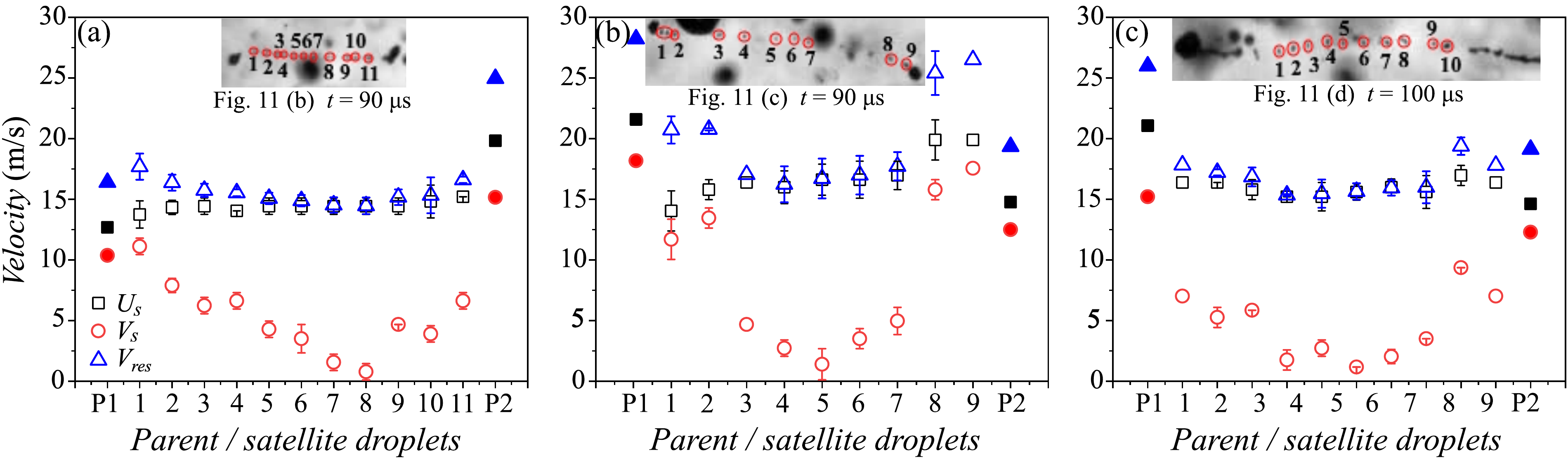}
	\end{center}
    \vspace{-10pt} 
   \caption{{Measured axial ($U_s$) and radial velocities ($V_s$) of parent (filled symbols) and satellite droplets (open symbols) for the \textit{stretching separation }shown in Figure \ref{fig:stretching}(b), (c), and (d), respectively.  $V_{res}$ is the resultant velocity. The number denotes the satellite droplets, annotated in the supplementary videos and shown in the graphs themselves.}} 
   \label{fig:stretching_vel}
\end{figure}
\end{nolinenumbers}

\begin{nolinenumbers}
\begin{figure}[!ht]
	\begin{center}		   
    \includegraphics[width=0.72\textwidth,keepaspectratio=true]{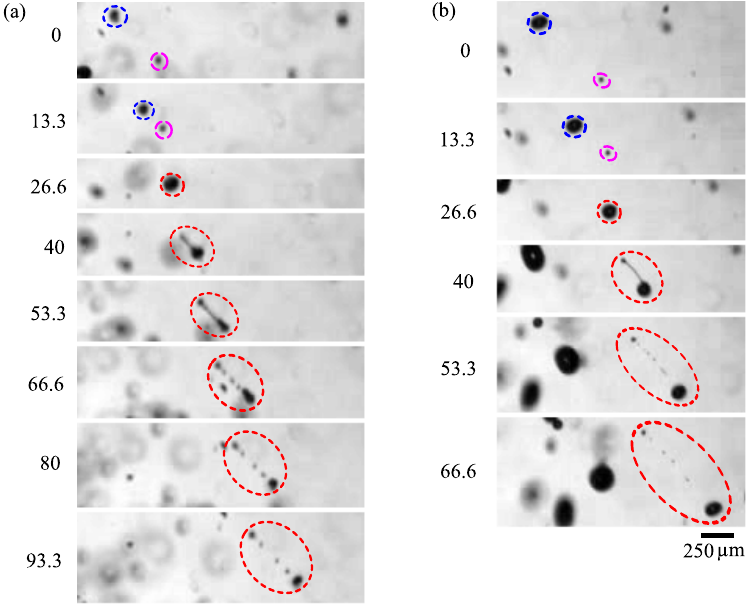}
	\end{center}
   \caption{Image sequences illustrating {eccentric binary droplet collision} in \textit{SS}. The cases shown correspond to \textit{stretching separation}:  (a)  $W\!e_s$ = 4.9 ($W\!e$ = 247, $v_r$ = 17.6 m/s, $d_L$ = 58 $\mu m$, $d_R$ = 88 $\mu m$, $W\!e_l$ = 2704), and  (b) $W\!e_s$ = 8.3 ($W\!e$ = 423, $v_r$ = 22.9 m/s, $d_L$ = 110 $\mu m$ , $d_R$ = 35 $\mu m$, and $W\!e_l$ = 2704). Time in $\mu$s is mentioned on the left. The scale bar for all cases is the same. The high-speed videos of the events are shown in \href{https://drive.google.com/file/d/1p8SwFVTR1zY-gqN6w69I-MdOu8_4jIMc/view?usp=sharing}{$MF-13a$} and \href{https://drive.google.com/file/d/1eOmB4YLDxr9cNLSD0O8n9WeQpZhxubCy/view?usp=sharing}{$MF-13b$}.}
   \label{fig:ss-single}
\end{figure}
\end{nolinenumbers}

In all cases of stretching separation, the fragmentation of the ligament leads to the formation of an array of satellites. The velocity of the satellites formed in Figure \ref{fig:stretching}(b) - (d) (see supplementary material) is sequentially illustrated in Figure \ref{fig:stretching_vel} from left to right. The analysis reveals that the variation of both $U_s$ and $V_s$ velocities exhibits near symmetry, which indicates a cohesive movement of the satellites as an array. $U_s$  of the satellites is observed to be nearly uniform in magnitude. The parabolic variation of $V_s$ across all cases is particularly notable. Droplets near the collision axis have minimal radial velocity, indicating that the liquid segment in this region remains stationary in the radial direction. Because of this parabolic velocity profile, most satellite droplets travel predominantly in the axial direction.

As previously mentioned, instances of stretching separation are also observed in \textit{SS}, although they occur less frequently. Two examples of such intra-spray collisions are illustrated in Figure \ref{fig:ss-single}. Since the colliding droplets in \textit{SS} move in the same direction, a faster-moving droplet (typically larger) may approach a slower one and collide with high eccentricity, leading to the formation of an array of satellite droplets. Unlike in \textit{CS}, where satellite droplets often exhibit dominant axial motion (Fig. \ref{fig:stretching_vel}), the satellite droplets here follow a trajectory similar to those of the parent droplets, as seen in frames after $t = $ 26.6 $\mu s$ in Figure \ref{fig:ss-single}. 

\clearpage

\hspace{-0.55cm}\underline{\textit{Stretching with digitations}}
\begin{nolinenumbers}
\begin{figure}[!ht]
	\begin{center}		   
    \includegraphics[width=1\textwidth,keepaspectratio=true]{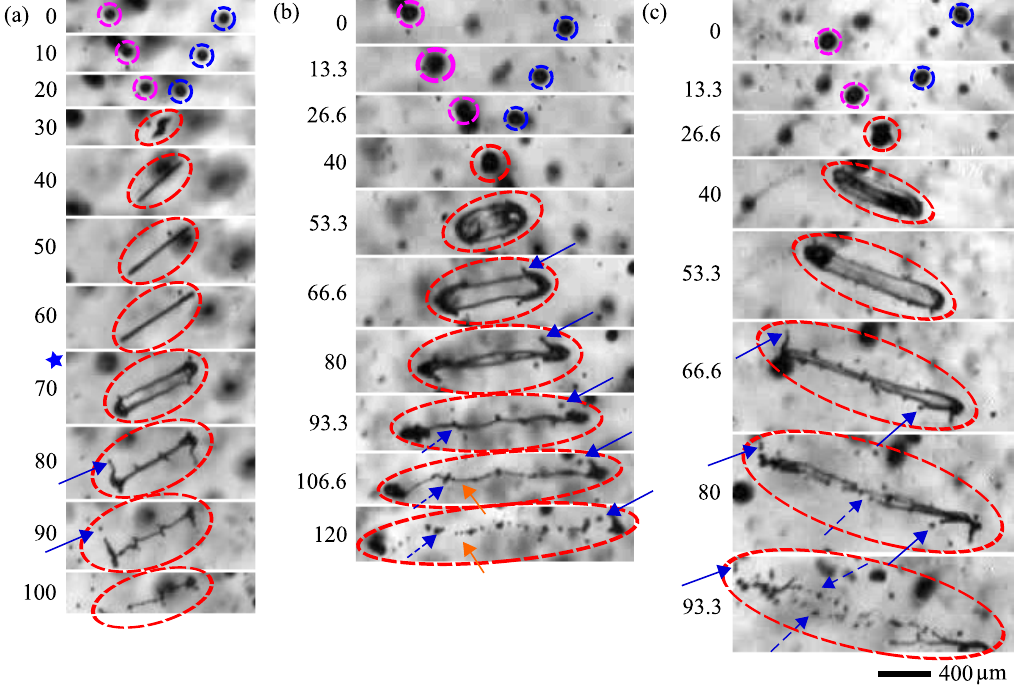}
	\end{center}
    \vspace{-10pt} 
   \caption{Image sequences illustrating \textit{stretching with digitations} from off-center binary droplet collisions with mid-range $B$ values in $CS$. The cases shown correspond to:  (a) $W\!e_s$ = 9.7 ($W\!e$ = 430, $v_r$ = 22.2 m/s, $d_L$ = 63 $\mu m$ , $d_R$ = 70 $\mu m$, $B$ = 0.31, and $W\!e_l$ = 1896), (b) $W\!e_s$ = 24.5 ($W\!e$ = 1176, $v_r$ = 29 m/s, $d_L$ = 105 $\mu m$, $d_R$ = 104 $\mu m$,  and $W\!e_l$ = 2704), and (c) $W\!e_s$ = 36.7 ($W\!e$ = 1704, $v_r$ = 36 m/s, $d_L$ = 121 $\mu m$, $d_R$ = 98 $\mu m$, $B$ = 0.54, and $W\!e_l$ = 2704). Time in $\mu$s is mentioned on the left. The scale bar for all cases is the same. The high-speed videos of the events are shown in \href{https://drive.google.com/file/d/1xs8bhVJEewVBiyg3BWYxXqU3x0NSFeh4/view?usp=sharing}{$MF-14a$}, \href{https://drive.google.com/file/d/1kZ3DByKRy4sk4iNf2-Lju2jR7-UQAkeC/view?usp=sharing}{$MF-14b$}, and \href{https://drive.google.com/file/d/1hjCg8On1_O_kg5Eqh47dFqTuHY5gN9kt/view?usp=sharing}{$MF-14c$}.}
   \label{fig:streching_with_digi}
\end{figure}
\end{nolinenumbers}

Figure \ref{fig:streching_with_digi}(a) illustrates an image sequence for the off-center collision of droplets, which leads to the formation of a symmetric ‘S’ shaped temporary coalesced structure at $t =  30\ \mu$s, which means that it is nearly an in-plane collision and calculation yields $B$ = 0.31. Under such mid-range of $B$, the phenomenon of \textit{stretching with digitations} is observed \citep{roth2007droplet}. The temporary coalesced structure is stretched along the direction of relative velocity to form a planar shape. At $t =  70\ \mu$s, the structure rotates and reveals the stretching lamella bounded by the TC rim with two thick round ends.  With further stretching, the rim collapses into a ligament with prominent nodes featuring long fingers at the extremes (formed by the collapsing of thicker ends), resembling the whiskers of a catfish. Smaller nodes along the ligament are also visible, transforming into small fingers ($t$ =  80 – 90 $\mu s$) which are nearly perpendicular to the ligament. Here, the transformation of the rim into ligament is similar to the coalescence of rims between two expanding holes on planar films, which creates fingers perpendicular to the rim (see Figure 1b in \cite{Nel2020}). A small satellite droplet is expelled from the left whisker as shown at $t$ =  80 – 90 $\mu s$, indicated by the arrow. Subsequently, the ligament, along with its smaller fingers, gives rise to more number of smaller satellites, while the end nodes contribute to the formation of larger satellite droplets (partially visible at $t$ = 100 $\mu s$).

In Figure \ref{fig:streching_with_digi}(b), the phenomenon is depicted at comparatively higher inertia with $W\!e_s$ = 24.5 from a different perspective as the droplet collides off-plane. A $t$ =  53.3 $\mu s$ an elliptical lamella with TC rim is formed. The lamella is stretched as shown at $t$ =  66.6 $\mu s$, having round ends with whiskers. At $t$ = 80 $\mu s$, similar knots previously seen in Figure \ref{fig:stretching}(d) are observed. At \( t = 93.3\,\mu\text{s} \), the rim collapses into a wavy ligament,   { closely resembling the structure reported in Figure~9a of}  \citet{Lo2025}. These observations point out that the formation of ligament is due to the twisting of the rim. The ejection of satellite droplets from the right whisker is visible at $t$ =  66.6 to 120 $\mu s$ (marked by arrows). The ligament exhibits numerous smaller nodes, one of which is highlighted by the dotted arrow at $t = $ 93.3 $\mu s$, which grow and transform into satellite droplets as indicated by the dotted arrow at $t = $ 106.6 and 120 $\mu s$. The ligament between any two nodes breaks into much smaller droplets, as shown by the orange arrow ($t = $ 106.6 and 120 $\mu s$). Additionally, the end nodes transform larger satellite droplets.

At very high inertia, as illustrated in Figure \ref{fig:streching_with_digi}(c) with $W\!e_s$ = 36.7 and $B$ = 0.54 (at $t = $ 26.6 $\mu s$, the side view of the temporary coalesced structure is clearly observed), there is a notable shift in the post-collision dynamics of ligament formation. By $t = $ 80 $\mu s$, the rim of the elliptical lamella breaks from the segment closure to the collision axis (indicated by a dotted arrow) and does not collapse into a single ligament. This behavior is attributed to thinning the rim’s portion near the collision axis under high inertial stretching, leading to its preferential breakup, as previously discussed in the context of high-inertia stretching separation. Each portion of the broken elliptical rim forms an array of droplets (two dotted arrows at $t = $ 93.3 $\mu s$). Moreover, the mechanism of end pinching of satellite droplets from end whiskers remains the same at high $W\!e_s$ (shown by solid arrows at $t = $ 66.6 to 80 $\mu s$). This regime results in a larger number of satellite droplets compared to the case of stretching separation as previously reported by \cite{roth2007droplet}. 
In some cases, the rim starts twisting but breaks before converting into the wavy ligament at high $W\!e_s$. Future off-center binary droplet collision studies should aim to elucidate the formation of ligaments,  whether it occurs through coalescence or twisting of the rim,  and their preferential breakup, as this process governs the size and velocity distribution of the resulting satellite droplets.

Irrespective of the intensity of inertial force in this regime, the stretched ligament forms a similar array of droplets as observed in the stretching separation (Fig. \ref{fig:stretching}). Consequently, the satellites' $U_s$ and $V_s$ trends are similar, with axial motion remaining dominant. For instance, Figure \ref{fig:strect_digi_vel}  illustrates the velocity variation of satellites formed in Figure \ref{fig:streching_with_digi}(b), where most exhibit dominant axial motion.	

\begin{nolinenumbers}
\begin{figure}[!ht]
	\begin{center}	     \includegraphics[width=0.47\textwidth,keepaspectratio=true]{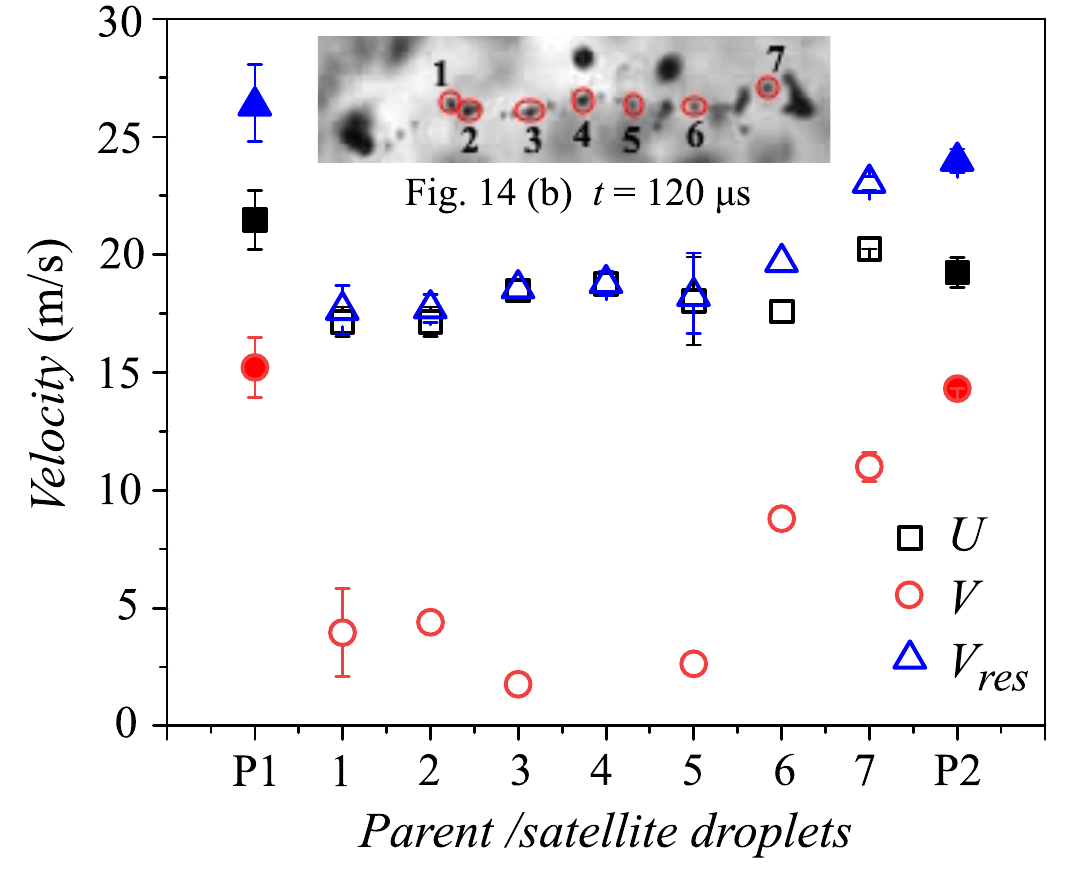}
	\end{center}
    \vspace{-10pt} 
   \caption{{Measured axial ($U_s$) and radial velocities ($V_s$) of parent (filled symbols) and satellite droplets (open symbols) for the \textit{stretching with digitations} shown in Figure \ref{fig:streching_with_digi}(b).} }
   \label{fig:strect_digi_vel}
\end{figure}
\end{nolinenumbers}

\subsubsection{\textbf{Multi-droplet collision}}\label{sec:multi}

\hspace{0.65cm}Previous discussions on head-on and off-center binary droplet collisions described the fundamental collision phenomena. However, it is imperative to delve into multi-droplet collisions, given their high prevalence, as illustrated in Figure~\ref{fig:collision_freq}. While a few studies have examined simultaneous ternary droplet collisions \citep{Hinterbichler2015,Yu2023} and provided intuitive insights, multi-droplet collisions in complex spray environments are more likely to occur sequentially. For instance, one or more droplets may impact a transiently coalesced structure formed by a prior binary collision. This sequential collision leads to the formation of complex liquid morphology that is significantly more difficult to characterize than binary collisions. However, a universal trend emerges: collision involving binary coalesced structure(s) and droplets leads to the formation of either a stretched ligament or an asymmetric lamella bounded by the TC rim, which subsequently collapses into a ligament. Figure \ref{fig:multi_collision}(a) exemplifies this phenomenon, showcasing a cascade of multiple stretching separation events occurring at $t = $ 0, 30, and 70 $\mu s$ resulting in the formation of ligaments at $t = $ 20, 50, and 90 (indicated by red, blue, and pink solid arrows). The ligaments ultimately result in arrays of droplets as seen at $t = $ 30, 90, and 100 $\mu s$ (indicated by red, blue, and pink dotted arrows).

\begin{nolinenumbers}
\begin{figure}[!ht]
	\begin{center}		   
    \includegraphics[width=\textwidth,keepaspectratio=true]{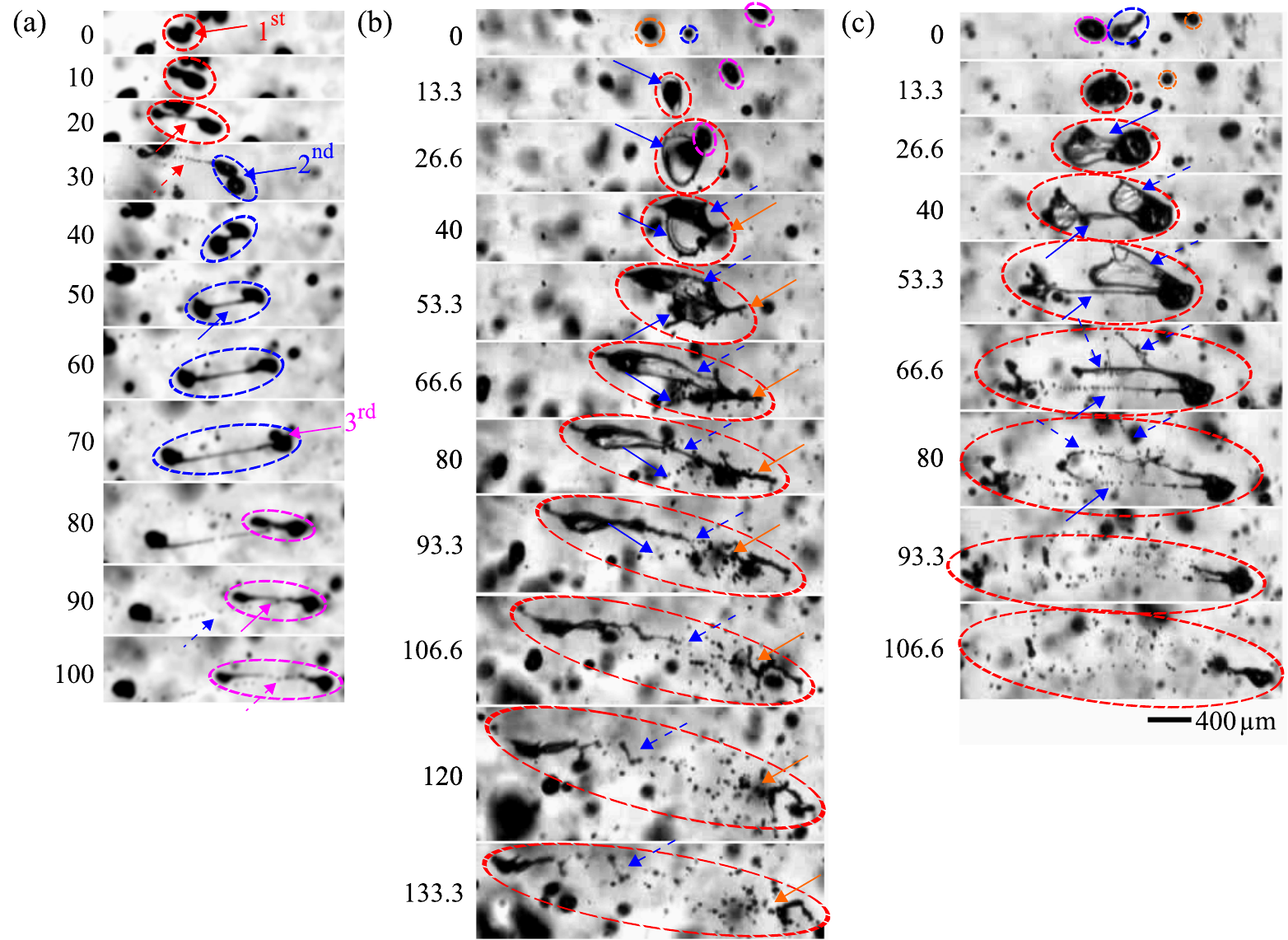}
	\end{center}
    \vspace{-10pt} 
   \caption{Temporal image sequences of multi-droplet collisions in $CS$ at $W\!e_l$ = 1896. The cases shown correspond to: (a) Multi-stretching separation, (b) triple droplet collision, and (c) droplet collision among droplets and a temporary coalesced liquid structure. Time in $\mu$s is mentioned on the left. The scale bar for all cases is the same. The high-speed videos of the events are shown in \href{https://drive.google.com/file/d/1EEHeZoEWr7TjECKlPKMl0mBUmhlFBnsW/view?usp=sharing}{$MF-16a$}, \href{https://drive.google.com/file/d/1nB55_Tq9dXxNWLsghGyp2r0Pg-ISxXtz/view?usp=sharing}{$MF-16b$}, and \href{https://drive.google.com/file/d/1fsu3kB2ClRHvzqS4EfWSSNKfx5sh8nBz/view?usp=sharing}{$MF-16c$}. }
   \label{fig:multi_collision}
\end{figure}
\end{nolinenumbers}

Figure \ref{fig:multi_collision}(b) captures the collision of two droplets encircled by {orange}  and blue ellipses at $t = $ 0 $\mu s$, resulting in the formation of a lamella depicted at $t = $ 13.3 - 26.6 $\mu s$ (indicated by solid blue arrow). This temporary coalesced structure collides with an incoming droplet seen at $t = $ 0 - 26.6 $\mu s$ (encircled by {pink} ellipse), leading to the formation of another asymmetric lamella seen at $t = $ 40 - 53.3 $\mu s$ (indicated by blue dashed arrow). The newly formed lamella progressively stretches, transitioning into a ligament (blue dashed arrow at $t = $ 66.6 - 80 $\mu s$), and initiating breakdown from $t = $ 93.3 $\mu s$, yielding numerous smaller satellite droplets at $t = $ 93.3 – 133.3 $\mu s$ (indicated by blue dashed arrow). Concurrently, the primary lamella also collapses into a ligament (indicated by blue solid arrow at $t = $ 53.3 - 66.6 $\mu s$), producing satellite droplets (indicated by blue solid arrow at $t = $ 80 – 93.3 $\mu s$). Moreover, the ligament formed at extreme right at $t = $ 40 $\mu s$  (indicated by orange solid arrow), undergoes stretching ($t = $ 40 – 80 $\mu s$) and eventual fragmentation and yields additional satellite droplets (orange solid arrow at $t = $ 93 - 133 $\mu s$).

Figure \ref{fig:multi_collision} (c) provides another example supporting the universality of stretched ligament formation in multi-droplet collision. At $t = $ 13.3 $\mu s$, a droplet (indicated by pink ellipse at $t = $ 0 $\mu s$) collides with a liquid structure (indicated by blue ellipse at $t = $ 0 $\mu s$), which seems to be a temporary coalesced structure formed from a binary collision. This collision forms a liquid structure with an asymmetric lamella at $t = $ 26.6 $\mu s$ (indicated by blue solid arrow). A subsequent droplet (indicated by orange ellipse at $t = $ 0 – 13.3 $\mu s$) collides with this liquid structure, resulting in the formation of another asymmetric lamella at $t = $ 40 $\mu s$ (indicated by blue dotted arrow). Both lamella stretches and collapse into ligaments at $t = $ 40 – 53.3 $\mu s$ (indicated by blue solid arrows) and $t = $ 53.3 - 66.6 $\mu s$ (indicated by blue dotted arrows), respectively. Eventually, these ligaments break at $t = $ 66.6 $\mu s$ (indicated by blue solid arrows) and 80 $\mu s$ (indicated by blue dotted arrows), forming a numerous array of satellite droplets as seen in later frames. As explained in \textit{stretching separation} and \textit{stretching with digitation} outcomes, most droplets formed from stretched ligaments have dominant axial motion. Therefore, the satellite droplets have similar characteristics in the case of multi-droplet collision. The propensity of satellite droplets, arising from both binary and multi-droplet collisions, to follow predominantly vertical trajectories leads to altered momentum distribution in \textit{CS}. 

\subsection{Size and velocity characteristics}

\hspace{0.65cm}To further quantify the effect of collisions on droplet dynamics, joint probability density functions ($J\!P\!D\!F\!s$) of droplet size and velocity components are constructed from the PDI measurements. These plots offer a statistical representation of how droplet size correlates with velocity in both axial and radial directions, thereby capturing the momentum redistribution induced by collision-driven fragmentation. The color of each point in a $J\!P\!D\!F$ represents the probability of the droplet with velocity components \( U \) (axial) and \( V \)(radial). The $J\!P\!D\!F\!s$ are generated using a MATLAB code \citep{nils2021scatter} that applies kernel smoothing to estimate the underlying probability densities. Figure~\ref{fig:fig_D_vs_U_pdi} presents the velocity--size correlations at \( z/G = 0.6 \) for \textit{SS} (subplots a--b) and \textit{CS} (subplots c--d), obtained from PDI at $W\!e_l$ = 1896.  {For \textit{SS}}, considering only the top 10\% of the data at $z/G = 0.6$, for $U$ velocity peak, droplets predominantly fall within $d = 50$ - $75~\mu$m and exhibit velocities of $U = 14$ - $17~\mathrm{m/s}$, 
whereas for $V$, the dominant size range is $d = 40$ - $66~\mu$m with corresponding velocities of $V = 9$ - $11~\mathrm{m/s}$. While the size ranges differ slightly, the diametric hot-cores for the $U$ and $V$ components overlap substantially, indicating that a common droplet-size band governs both $U$ and $V$ velocity peaks. 
These droplets are detected at the boundary of \textit{SS}, formed via conical sheet breakup, and thus, carry dominant radial velocity. In the case of \textit{CS}, the $J\!P\!D\!F\!s$ at \( z/G = 0.6 \) exhibit a distinct shift in the dominant droplet population toward smaller sizes, primarily in the range of $28$ – $47~\mu$m, accompanied by a modest reduction in axial velocity, now concentrated around $12$–$15~\mathrm{m/s}$. 
The $V$–$d$ $J\!P\!D\!F$ are nearly symmetric about the vertical axis, with $V$  spanning both positive and negative values, ranging from $-5$ to $5~\mathrm{m/s}$, and corresponding droplet sizes clustered around $23$–$42~\mu$m, closely matching those observed for the $U$ component. 
This reinforces the earlier observation that a similar droplet-size band governs both components. Interestingly, $J\!P\!D\!F\!s$ at $We_l = 2704$  (subplot (e-h) in Figure~\ref{fig:fig_D_vs_U_pdi}) display similar characteristics, further supporting the trends.
It is evident that velocity--size correlations differ significantly between \textit{SS} and \textit{CS}. This variation arises not only from the geometric overlap of the two sprays but also from the underlying collision dynamics. In a scenario of pure geometric overlap without actual droplet collision, the $J\!P\!D\!F\!s$ of \( U\text{-}D \) for \textit{SS} and \textit{CS} would be expected to remain identical, as the velocity \( U \) is directed similarly in both cases and, without collisions, droplet size \( D \) should remain unaffected. While droplets with sizes and velocities similar to those in \textit{SS} still appear in \textit{CS}, their occurrence is less intense, as inter-spray collisions deplete the population of large parent droplets and lead to the formation of finer droplets.
High-speed imaging, supported by quantitative data in Figures~\ref{fig:vel_splashing}, \ref{fig:stretching_vel}, and \ref{fig:strect_digi_vel}, confirms that the smaller droplets originate from collisions. The resulting smaller satellite droplets typically exhibit slightly lower axial velocities than their parent droplets (see solid symbols in Fig.~\ref{fig:vel_splashing}, \ref{fig:stretching_vel}, and \ref{fig:strect_digi_vel}) but remain strongly axially dominated, with the radial component remaining comparatively small. In contrast, the stretching separation in \textit{SS} leads to the formation of satellite droplets that closely follow the trajectory of their parent (Fig. \ref{fig:ss-single}), resulting in minimal alteration of the overall momentum distribution.

\begin{nolinenumbers}
\begin{figure}[H]
    \begin{center}		   
    \includegraphics[width=\textwidth,keepaspectratio=true]{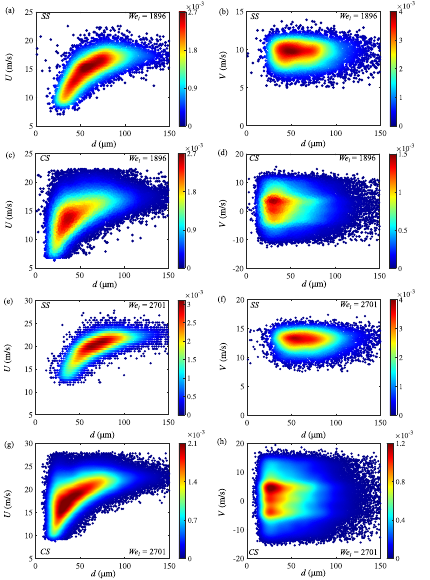}
	\end{center}
    \vspace{-10pt} 
   \caption{$J\!P\!D\!F\!s$ of droplet diameter ($d$) and velocity components ($U$ and $V$) at $z/G = 0.6$ for $SS$ and $CS$ cases for two $W\!e_l$. (a–d): $We_l = 1896$; (e–h): $We_l = 2704$. For each $W\!e_l$, (a,c,e,g): $U$–$d$ plots; (b,d,f,h): $V$–$d$ plots. $SS$ appear in (a–b,e–f), and $CS$ in (c–d,g–h).}
   \label{fig:fig_D_vs_U_pdi}
\end{figure}
\end{nolinenumbers}

\clearpage
\hspace{-0.65 cm}

\begin{nolinenumbers}
 \begin{figure}[!ht]
      \begin{center}		   
     \includegraphics[width=0.85\textwidth,keepaspectratio=true]{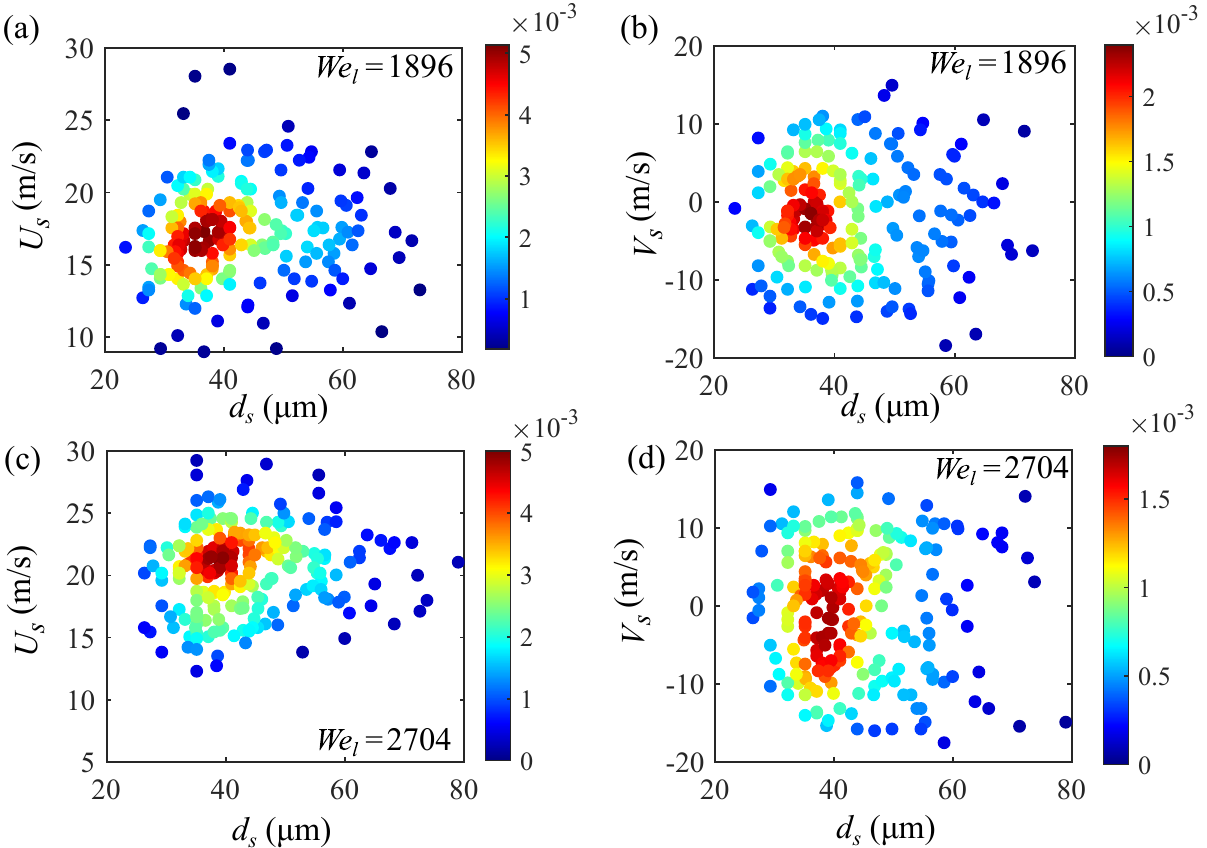}
 	\end{center}
    \vspace{-10pt} 
 	\caption{$J\!P\!D\!F\!s$ of size–velocity correlations for satellite droplets observed in high-speed videos of \textit{CS} at different $W\!e_l$.  (a–b): $We_l = 1896$; (c–d): $We_l = 2704$. (a,c): $U_s$ vs $d_s$; (b,d): $V_s$ vs $d_s$.}
   \label{fig:fig_D_vs_u_s-v_s_images}
 \end{figure}
 \end{nolinenumbers}
 
To support the above reasoning, the diameter ($d_s$) and axial ($U_s$) and radial ($V_s$) velocities of satellite droplets in \textit{CS} are extracted through frame-by-frame analysis of high-speed video sequences, as presented earlier. For each $W\!e_l$, approximately 200 satellite droplets, formed during different collision events, are manually tracked to determine their size and velocity. Due to the resolution limits of the imaging system, finer satellites with $d_s < 20~\mu$m are not measured. Figure~\ref{fig:fig_D_vs_u_s-v_s_images}(a) and (b) present $U_s$ - $d_s$ and $V_s$ - $d_s$ $J\!P\!D\!F\!s$, respectively, for $We_l = 1896$. Comparing these plots with Figure~\ref{fig:fig_D_vs_U_pdi} (c) and (d), it is evident that the most dominant regions occupy similar size and velocity ranges. This close match supports the conclusion that the dominant droplet class in \textit{CS} primarily consists of satellite droplets. A similar trend is observed for $We_l = 2704$, as seen by comparing Figure~\ref{fig:fig_D_vs_u_s-v_s_images}(c) and (d) with Figure~\ref{fig:fig_D_vs_U_pdi}(e) and (f), respectively. Notably, the work of \citet{Ghosh2025} on \textit{GCSC} sprays highlights that satellite droplets are primarily generated via stretching separation, which also emerges as one of the dominant mechanisms in the present study. Due to their lower inertia, these satellites quickly adjust to the axial air velocity imparted by the central air jet of \textit{GCSC}, increasing their axial velocity. However, as shown in the current study, satellite droplets inherently possess a tendency to move in the axial direction. Aerodynamic forces in such configurations further amplify this intrinsic axial motion. It is also worth noting that the present study does not observe bag-breakup-type fragmentation, reported in \citet{Ghosh2025}.

\subsubsection{\textbf{Collision cascade induced by satellite droplets}}
    
\begin{nolinenumbers}
 \begin{figure}[!ht]
   \begin{center}		   
     \includegraphics[width=\textwidth,keepaspectratio=true]{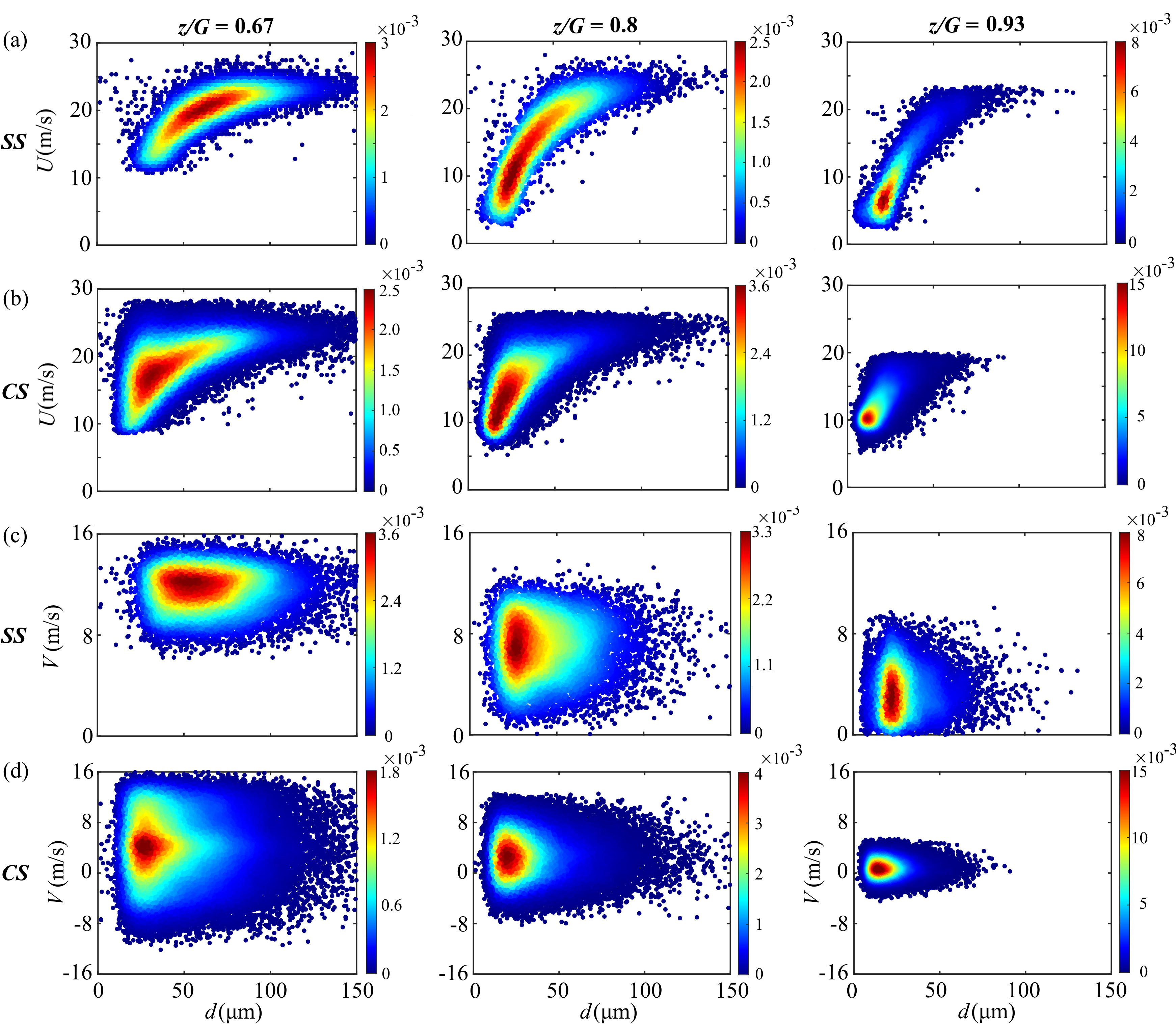}
 	\end{center}
    \vspace{-10pt} 
   \caption{$J\!P\!D\!F\!s$ of droplet diameter ($d$) with axial ($U$) and radial ($V$) velocity components for $SS$ and $CS$ at $We_l = 2704$ over the downstream range $z/G = 0.66$--0.93. Panels (a) and (c) correspond to $U$--$d$ and $V$--$d$ distributions for $SS$, respectively, while (b) and (d) show the corresponding results for $CS$.}
   \label{fig:jpdf_cascade}
\end{figure}
\end{nolinenumbers}
\hspace{0.65cm} 

 \hspace{0.65cm} \textcolor{black}{The discussion so far has shown that, at the onset of the interaction region, the smaller dominant droplets in 
$CS$ exhibits predominantly axial motion, corresponding to satellites generated by high-energy collisions. Following the onset analysis, the downstream evolution of droplet size and velocity is examined through the $J\!P\!D\!F\!s$ of $U$–$d$ and $V-d$ obtained from PDI presented in Figure \ref{fig:jpdf_cascade} for $SS$ and $CS$ at $W\!e_l$ = 2704. As the probe volume moves downstream in $SS$, it progressively enters the hollow region, leading to a steady reduction in size and axial velocity of the dominant class (Fig. \ref{fig:jpdf_cascade} (a)). Meanwhile, as shown in Figure \ref{fig:jpdf_cascade} (b), for $CS$ at $z/G = 0.66$, the dominant droplets are smaller with a slight reduction in $U$, likely because the dominant satellite droplets have somewhat lower axial velocities than their parent droplets (Figs. \ref{fig:vel_splashing}, \ref{fig:stretching_vel}, and \ref{fig:strect_digi_vel}). However, since these satellites possess stronger streamwise motion than those in $SS$, they travel farther downstream with only a mild decay in $U$, leading to a higher axial velocity at the end of the interaction region ($z/G=0.93$). The $V-d$ $J\!P\!D\!F\!s$  in Figure \ref{fig:jpdf_cascade} (c) and (d) further demonstrate that the droplet motion remains predominantly axial downstream of the interaction onset. The $V$ velocity of dominant droplets in $CS$ remains lower than that in $SS$ for all $z/G$. By the end of the interaction ($z/G = 0.93$), the $CS$ droplets become increasingly concentrated around $V \approx 0$, indicating the predominance of axial motion. Interestingly, as previously shown in Figure \ref{fig:droplet_dis}, the size of the dominant droplets in $CS$ continues to decrease, accompanied by an apparent rise in local number density (see the color bars for $z/G = 0.93$ in Figs.~\ref{fig:jpdf_cascade} (b) and (d)). The increased number density of axially oriented smaller droplets indicates continued high-energy collisions downstream. These correlated changes in size, number density, and velocity components provide strong statistical evidence of a collision-driven cascade process. Such a cascade phenomenon is plausible, as evidenced by the multi-droplet collisions in Figure \ref{fig:multi_collision}.  In addition to parent–parent interactions, axial moving satellite droplets can collide with other satellites or parents, driving further breakup and generating even smaller droplets. These observations demonstrate that the $CS$ configuration not only modifies the mass and momentum distribution but also fundamentally alters the downstream breakup dynamics, leading to enhanced atomization and a denser population of fine droplets through collision cascade driven by satellite droplets.}

\subsection{Ligament-mediated breakup and droplet-size distribution}
\begin{nolinenumbers}
\begin{figure}[!ht]
   \begin{center}		   
     \includegraphics[width=0.70\textwidth,keepaspectratio=true]{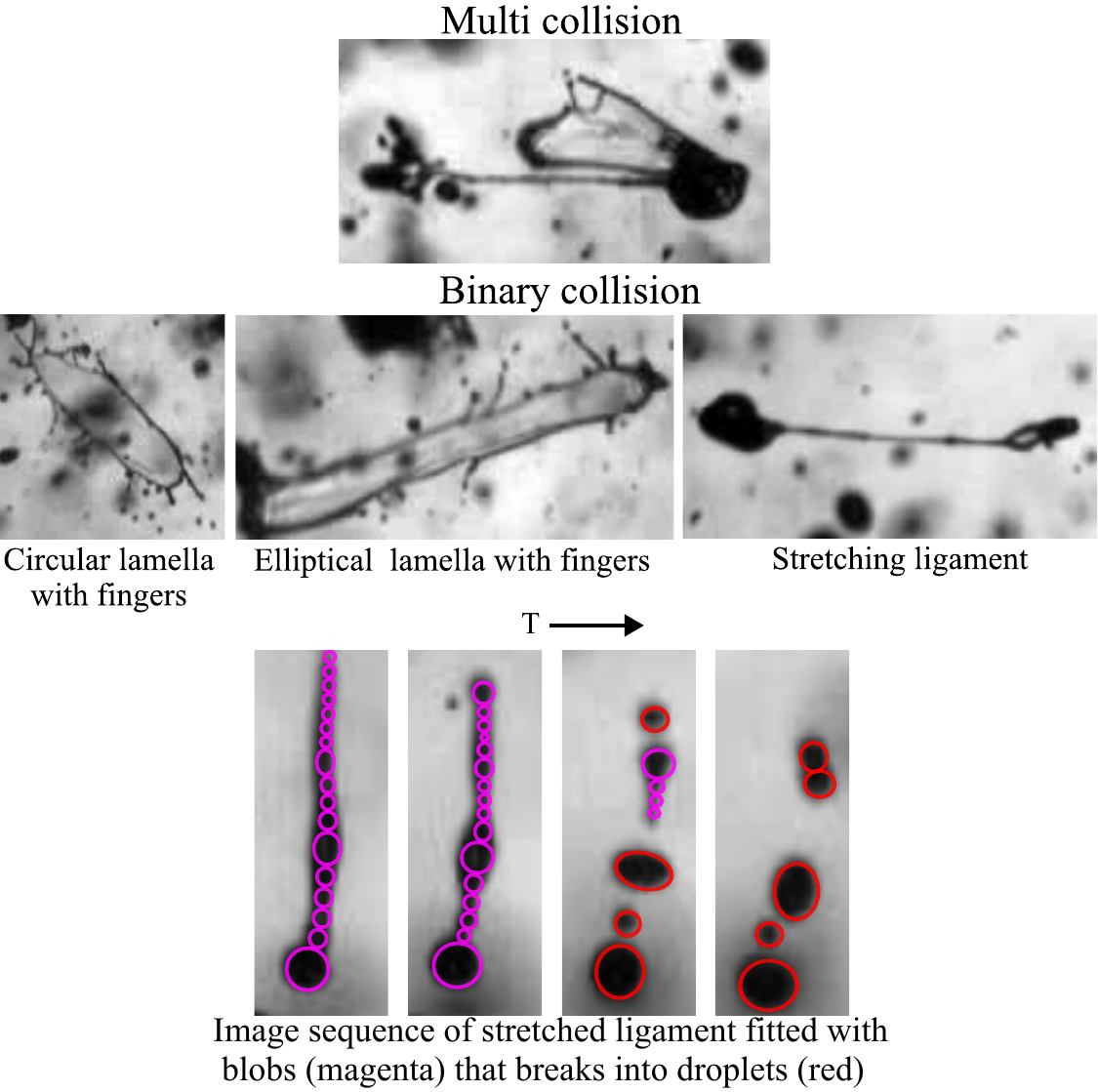}
 	\end{center}
    \vspace{-10pt} 
  \caption{Images representing 
  multi-droplet collisions (top row) and binary collision events (middle row). Ligament‐mediated fragmentation underpins every collision and can be resolved into individual blobs colored in magenta (bottom row), whose size distribution is well characterized by a gamma distribution. The final droplets are marked by red color. The time ($T$) is increasing from left to right. }

   \label{fig:gamma_blobs}
\end{figure}
\end{nolinenumbers}

 \hspace{0.65cm}The preceding observations establish that droplet collisions in $CS$ lead to the formation of stretched ligaments, fingers, and lamella, which subsequently fragment into satellite droplets. These secondary structures play a dominant role in shaping the droplet size distribution within the interacting region. Figure \ref{fig:gamma_blobs} shows the representative structures formed in the multi and binary droplet collisions. Although the intermediate liquid structure differs in each of these cases, their final form before disintegration into droplets is that of a ligament. An isolated ligament can be envisioned as a train of blobs, as shown by the schematic in the bottom row in Figure \ref{fig:gamma_blobs} (first and second column). \textcolor{black}{These blobs do not represent final detached droplets but rather the instantaneous
liquid bulges that develop along a stretching ligament, consistent with the kinetic model of \citet{Marmottant}. In that framework (Eq. 11 and Fig. 18 in \citet{Marmottant}), successive rearrangements and coalescence events
progressively reduce the number of blobs while the ligament roughness increases and eventually saturates as
breakup is approached. The smaller blobs in the thinnest region, therefore, correspond to the transient, finer blobs
that arise during this evolution and eventually merge or pinch off into droplets, marked by red blobs in the third
and fourth columns.}  Such a transformation appeals favorably to ligament-mediated droplet formation and strongly supports the use of the gamma distribution to describe the ensuing droplet statistics \citep{marmottant2004spray}.
 The gamma distribution, $P_b$, in such cases is described as follows,
\begin{equation}
P_b \left( n, x = \frac{d}{\langle d \rangle} \right) = \frac{n^n}{\Gamma(n)} x^{n-1} e^{-nx}.
\end{equation}
Here, the parameter $n$ characterizes the degree of ligament corrugation, $x = \frac{d}{\langle d \rangle}$ is the rescaled droplet diameter, $\langle d \rangle$ denotes the average droplet diameter, and $\Gamma$ represents the gamma function.
 Building on their investigation of droplet impact on solid surfaces with the same size as the droplet diameter, \citet{VILLERMAUX2011} showed that the rescaled droplet size distribution resulting from the breakup of ligaments of varying sizes is well described by a two-parameter compound gamma distribution, given by
\begin{equation}\label{eq:compound_gamma}
P_{m,n} \left( x = \frac{d}{\langle d \rangle} \right) = \frac{2 (mn)}{\Gamma(m) \Gamma(n)} x^{\frac{m+n}{2} - 1} K_{m-n} \left( 2 \sqrt{mnx} \right).
\end{equation}
Here $m$ determines ligament size distribution and $K$ is the modified Bessel function of the second kind of order $m$ - $n$. For highly corrugated ligaments, the value of $n$ is between 4 to 6, reaching $n$ = $\infty$ for exceptionally smooth ligaments, while a higher value of $m$ indicates narrow ligament size distribution and vice–versa \citep{Kooij2018,Sijs2021,Vankeswaram2022}. Furthermore, \citet{Kooij2018} and \citet{Sijs2021} demonstrated that the compound gamma distribution effectively predicts the rescaled size distribution of droplets in poly-disperse sprays.
Following this approach, we fit the rescaled droplet–size distributions of $SS$ and $CS$ using Eq. \ref{eq:compound_gamma} to determine the best-fit parameters $m$ and $n$, in line with the methodology of \citet{Kooij2019}. The experimental probability density value for each size class, $PDF_i$, is computed using the approach outlined by \citet{Vankeswaram2022} and is expressed as,

\begin{equation}
PDF_i = \frac{n_i}{\sum n_i} \left( \frac{\langle d \rangle}{\updelta d} \right).
\end{equation}

\hspace{-0.65cm}Here $n_i$ is the number of droplets in $i^{th}$ size class, and $\updelta d$ is the width of the size class equal to 5 $\mu m$. Figure \ref{fig:gamma} illustrates the rescaled size distribution for $SS$ and $CS$ for two $W\!e_l$. 
In Figure \ref{fig:gamma}(a), for $SS$ at  $W\!e_l$ = 1896, the data collapse very well at the boundary of the spray, i.e., $z/G$ =  0.60 and 0.67, and are well represented by the gamma distribution \( P_{m=100, n=5} \) delineated by the solid red curve. The droplets at the spray's edge are formed from the breakup of ligaments formed from the conical liquid sheet, thus conforming to the gamma distribution. The values $m$ = 100 and $n$ = 5 used to fit the experimental data are similar to those determined using ligament images in prior studies for hollow cone spray with water as the experimental fluid \citep{Kooij2018}. The higher value of $m$ indicates that the single hollow cone spray exhibits similar ligament size characteristics. Note that in the case of \textit{SS}, with increasing \( z/G \) along the centroidal axis, the probe volume traverses from the spray boundary into the less dense interior of the hollow-cone spray. As it passes through the hollow region, deviations from the gamma distribution become evident, highlighting its reduced suitability in such zones. A similarly good match is observed for $SS$ at \( We_l = 2704 \) in Figure \ref{fig:gamma}(c), at \( z/G = 0.60 \) to 0.73. This suggests an expanded region of droplets originating from the conical sheet, toward the hollow zone,  where the gamma distribution remains applicable, primarily due to a change in the spray cone angle with an increase in $W\!e_l$.

\begin{nolinenumbers}
\begin{figure}[!ht]
	\begin{center}		   
    \includegraphics[width=0.825\textwidth,keepaspectratio=true]{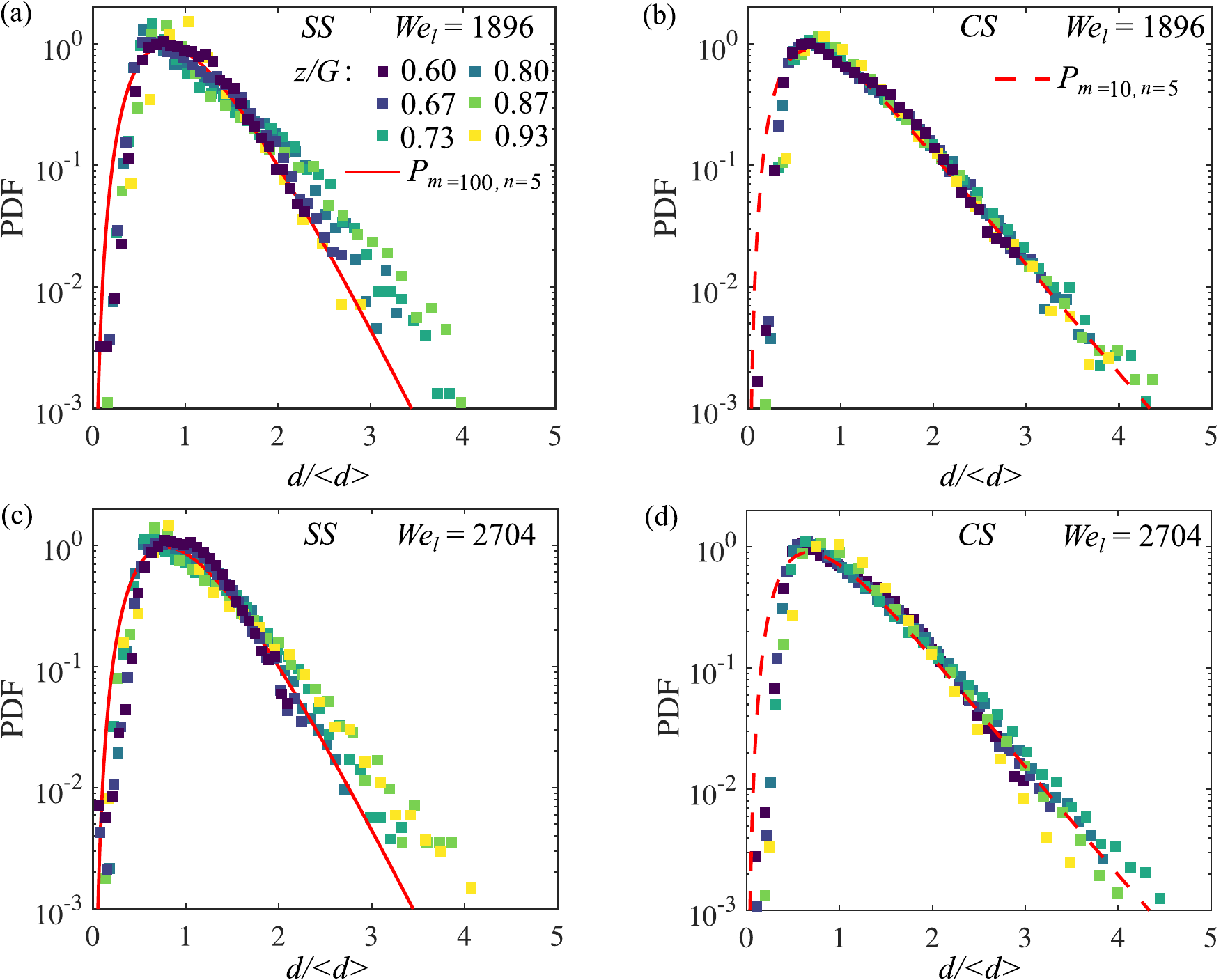}
	\end{center}
    \vspace{-10pt} 
   \caption{The PDF of droplet size distribution for $SS$ ((a) and (c)) and $CS$ ((b) and (d)) at different $z/G$ along the centroid axis for two $W\!e_l$. The dashed line corresponds to the theoretical prediction obtained using Eq. \ref{eq:compound_gamma}.}
   \label{fig:gamma}
\end{figure}
\end{nolinenumbers}

Figure \ref{fig:gamma}(b) and (d) illustrate the rescaled distribution for $CC$ for two $W\!e_l$. The data collapse very well across all $z/G$, and the distribution width increases compared to $SS$ because $\langle d \rangle$ decreases. Remarkably, for $CS$, using $m$ = 10 and $n$ = 5 yields an excellent match with the experimental data across all $z/G$ values. This agreement is attributed to the distinctive nature of satellite droplet formation from ligaments during collision events in $CS$. The decrease in the value of \( m \) from 100 to 10 provides a quantitative representation of the broader ligament size distribution observed during several different types of droplet collisions in  $CS$, in contrast to the narrower distributions resulting from the breakup of a conical sheet in $SS$. Moreover, the excellent fit of the gamma distribution even at higher $z/G$ suggests that these droplets correspond predominantly to satellite droplets. This further supports the argument that the increased number of smaller droplets at higher \( z/G \) results from enhanced cascade collisions, primarily driven by axially traveling satellite droplets formed at lower \( z/G \). However, it is essential to acknowledge that the gamma distribution slightly overestimates the population of very small-sized droplets, albeit successfully capturing the peak value.

\section{Conclusion and perspectives}
\vspace{-0.4cm}
\hspace{0.65cm}In this study, we examined the interacting region formed by mixing three identical hollow-cone sprays using PDI and microscopic high-speed backlight imaging. PDI measurements revealed a marked decrease in droplet size within the interaction region as the number of smaller droplets increased tremendously, which shifts the distribution towards the left. Imaging confirmed frequent high-energy collisions that led to the formation of numerous smaller droplets, leading to a decrease in local SMD. While earlier spray studies predominantly reported only stretching separation, our experiments demonstrate the presence of reflexive separation, fingering, splashing,  and stretching with digitations, phenomena previously seen only in highly controlled binary droplet collision studies. We present a taxonomy of these collision outcomes, including the evolution of transient coalesced structure, satellite formation, and their velocity signatures. Additionally, we report high-energy variants of known binary outcomes, featuring longer fingers during head-on collisions and twisting lamellae that rupture near the collision axis in off-center collisions. Beyond binary collisions, we observed sequential multi-droplet collisions, wherein a droplet impacts a transiently coalesced binary collision structure. These multi-collisions exhibit a universal outcome, forming either a stretched ligament or an elongated lamella that eventually transforms into a ligament. Tracking satellite droplets reveals that most of them have dominant axial motion, regardless of the collision type. The collisions alter the size–velocity correlations for $CS$, as the size of the dominant class decreases sharply, accompanied by a slight reduction in axial velocity but a pronounced decrease in radial velocity, as evidenced by $J\!P\!D\!F\!s$. Consequently, satellites formed at lower $z/G$ can further collide in cascade events at higher $z/G$, enhancing the shift toward finer droplets. Thereby shifting the size distribution peak towards finer droplets with increasing \( z/G \). The compound gamma distribution well predicts the rescaled droplet size distribution for $SS$ near the spray boundary. Interestingly, it also predicts the distribution for $CS$ for the whole interaction region investigated. The gamma distribution for $CS$ works because of the production of numerous satellite droplets from ligaments, which are the underlying fundamental structures in post-collision events (see the hierarchy in Figure \ref{fig:gamma_blobs}). The decrease in the value of $m$ for $CS$ highlights the broader spectrum of ligament sizes produced during droplet collision compared to those in the breakup of the conical liquid sheet in $SS$. 
 
{This study confirms that binary collisions remain relevant, although multiple droplet interactions are prevalent under practical spray conditions. Despite the macroscopically dense appearance of the spray, the spacing between droplets is far greater than their diameter, allowing individual binary interactions to occur. While the current analysis focused on a controlled configuration involving hollow-cone sprays, it is likely that both binary and multiple droplet collisions, along with the underlying velocity characteristics of satellites, may persist in more complex spray environments such as sea sprays. \textcolor{black}{Notably, binary collisions have been documented in mid-air interactions between raindrops \citep{Testik2017}, and similar collisions were observed near the pool surface in laboratory rainfall experiments \citep{Liu2024}. While multiple collisions may occur in regions of high droplet density.} Nonetheless, in natural and industrial settings, aerodynamic forces are expected to play a crucial role, as suggested by controlled studies such as those by \citet{hardalupas1996interaction} and \citet{Ghosh2025}. However, what is missing in the present study is a comprehensive regime map illustrating high-energy collision outcomes in the \( We \)–\( B \) space,  which remains in notable paucity compared to the low Weber number case.  Therefore, future research should focus on high-energy binary collisions under controlled yet representative conditions. From the authors’ perspective, the current work serves as a strong motivation toward that goal.}

\section{Acknowlegement}
\hspace{0.65cm}This work was carried out with the support of the National Center for Combustion Research and Development (NCCRD), Indian Institute of Science, India. The authors gratefully acknowledge Mr. R. Sakthikumar for his assistance in conducting the experiments. Additionally, the authors thank Mr. Vishal Singh, whose advice to use LDM proved to be a pivotal contribution to the direction of this study.

\bibliographystyle{abbrvnat_rev}
\bibliography{ref}   
\end{document}